\documentclass[sn-mathphys-num, iicol]{sn-jnl}
\usepackage{graphicx}
\usepackage{multirow}
\usepackage{amsmath,amssymb,amsfonts}
\usepackage{amsthm}
\usepackage{mathrsfs}
\usepackage[title]{appendix}
\usepackage{xcolor}
\usepackage{textcomp}
\usepackage{manyfoot}
\usepackage{booktabs}
\usepackage{algorithm}
\usepackage{algorithmicx}
\usepackage{algpseudocode}
\usepackage{listings}
\usepackage{siunitx}
\usepackage{xspace}
\usepackage{cleveref}
\usepackage{subcaption}
\usepackage[normalem]{ulem}
\usepackage{graphicx}
\usepackage{wasysym}
\usepackage{float}
\usepackage{comment}
\usepackage{soul}
\usepackage{lineno}
\usepackage[nohyperlinks, printonlyused, withpage, smaller]{acronym}

\theoremstyle{thmstyleone}%

\theoremstyle{thmstyletwo}%

\theoremstyle{thmstylethree}%

\newcommand{\sonoarrivatoqui}[1]{\textcolor{red}{\\ \%\%\% sono arrivato qui \%\%\% \\\\}}

\newcommand{\mainlens}{main lens\xspace}
\newcommand{\mla}{MLA\xspace}
\newcommand{\microlens}{micro-lens\xspace}

\newcommand{\neuralnetwork}{NN\xspace}

\newcommand{\SrSource}{$^{90}\text{Sr}$\xspace}
\newcommand{\photorig}{photon origin\xspace}
\newcommand{\ccOneMuZeroPiOneProt}{CC $1\mu 0\pi 1p$\xspace}
\newcommand{\ccOneMuZeroPiTwoProt}{CC $1\mu 0\pi 2p$\xspace}
\newcommand{\bce}{\ensuremath{BCE}\xspace}
\newcommand{\nuless}{\ensuremath{\nu}-less\xspace
\ensuremath{\beta\beta}\xspace}

\makeatletter
\renewcommand{\@title}{
  \begin{center}
    \LARGE
    An ultrafast plenoptic-camera system for high-resolution\\
    3D particle tracking in unsegmented scintillators
  \end{center}
}
\makeatother

\begin{document}

\author[1]{\fnm{Till} \sur{Dieminger}}
\author[1]{\fnm{Sa\'ul} \sur{Alonso-Monsalve}}
\author[1]{\fnm{Christoph} \sur{Alt}}

\author[2]{\fnm{Claudio} \sur{Bruschini}}
\author[1]{\fnm{Noemi} \sur{Bührer}}
\author[2]{\fnm{Edoardo} \sur{Charbon}}
\author[2]{\fnm{Kodai} \sur{Kaneyasu}}
\author[1]{\fnm{Tim} \sur{Weber}}
\author[1]{\fnm{Matthew} \sur{Franks}}

\author*[1]{\fnm{Davide} \sur{Sgalaberna}}\email{davide.sgalaberna@cern.ch}

\affil[1]{\orgdiv{IPA}, \orgname{ETH Zürich},
  \orgaddress{\street{Otto Stern Weg 5}, \city{Zurich},
\postcode{8093}, \state{Zurich}, \country{Switzerland}}}

\affil[2]{\orgdiv{Advanced Quantum Architecture Lab (AQUA)},
  \orgname{EPFL}, \orgaddress{\street{Rue de la Maladi\`ere},
\postcode{2000}, \city{Neuch\^atel}, \country{Switzerland}}}

\abstract{
Neutrino detectors, particle calorimeters and some dark matter detectors require dense and massive active materials. An extremely fine segmentation is desirable to achieve precise three-dimensional particle tracking. However, such systems introduce significant challenges in construction and demand a large number of readout electronics channels, leading to extremely high costs.
In this article, we propose an alternative approach to elementary particle detection that enables ultrafast three-dimensional high-resolution imaging in large volumes of unsegmented scintillator.
Enabling technologies are plenoptic systems and time-resolving single-photon avalanche diode array imaging sensors. Together, they enabled us, using a plenoptic camera, to reconstruct the origin of single photons in the scintillator.
A case study focused on neutrino detection demonstrates full event reconstruction with a spatial resolution of two hundred micrometres.
This work paves the way for a class of particle detectors whose capabilities should be further enhanced through future developments and expanded to Cherenkov light detection, medical imaging and neutron detection.
}

\keywords{particle detector, SPAD array sensor, organic scintillator,
plenoptic camera, 3D imaging, ray tracing, neutrino detector, deep learning.}
\maketitle

\section{Introduction}
\label{sec:intro}

Modern elementary particle detectors deployed at high-energy physics
experiments often have to cover huge volumes or surfaces and, at the
same time, provide a spatial resolution on the order of 100 $\mu$m
with a time resolution of a few hundred picoseconds or better.
Moreover, dense active materials are necessary for detecting weakly
interacting particles,
such as neutrinos
\cite{Hyper-Kamiokande:2025fci,Hyper-Kamiokande:2018ofw,
Abi:2020qib,Abe:2019vii,NOvA:2022see,MINOS:2008hdf}
and certain categories of dark matter
candidates~\cite{Zani:2024ybb,DarkSide-20k:2024yfq,Aprile:2020vtw,Aalbers:2022dzr},
as well as for electromagnetic and hadronic
calorimetry~\cite{calice,CALICE:2018ibt,2137105,CMSHCAL:2007zcq,CMS:2009cmd}.

Scintillator-based detectors constitute a suitable choice for active,
high-density detection systems, offering the potential for
sub-millimetre spatial resolution.
Organic scintillators emit blue or green isotropic light, typically
in the range of 8,000 to 13,000 photons per MeV loss, when traversed
by ionising radiation~\cite{Kolanoski:2020ksk}, with a decay time
down to about 1-2~ns~\cite{eljen-catalogue-2025,Artikov:2005mg}.
Advances in detector technology have enabled 3D segmentation down to
1~cm with sub-nanosecond timing in tonne-scale scintillator
volumes with wavelength shifting (WLS) fibre readout~\cite{Sgalaberna:2017khy,sfgd-testbeam-cern,Mineev:2018ekk,Alekseev:2022jki,Boyarintsev_2021}.
R\&D on 3D printing of plastic scintillator is also
underway~\cite{son2018improved, KIM20202910,
Berns:2020ehg,Berns_2022,Sibilieva:2022ket,KAPLON2022106864,Weber:2023joy}.
Another notable scintillator-based technology, currently under development, aims to get rid of the geometrical segmentation by deploying organic scintillating materials with a short optical scattering length to stochastically contain the visible photons within small localised volumes \cite{LiquidO:2019mxd}.
These detectors, typically read out by Silicon Photomultipliers
(SiPMs)~\cite{hamamatsu:mppc}, are also suitable for precise
time-of-flight measurements of both charged
particles~\cite{Korzenev:2021mny} and
neutrons~\cite{Agarwal:2022kiv,Gwon:2022bix}.
In the form of thin scintillating fibres, spatial resolutions below
$100~\mu\text{m}$ can be achieved~\cite{Joram:2015ymp,Papa:2023uqv}.
However, the main drawback lies in the challenging and costly
construction, due to the very large number of channels required,
particularly when targeting tonne-scale detectors.
However, the main drawback lies in the challenging and costly construction due to the very large number of analog readout channels required, each needing to be instrumented with corresponding digitisation electronics, particularly when targeting tonne-scale detectors.

In recent years, a new class of photosensors has seen tremendous
improvements: Single-Photon Avalanche Diode (SPAD) arrays.
These arrays consist of multiple photosensitive diodes manufactured
in CMOS technology, which are independently read out to provide
single-photon images with sub-nanosecond time resolution.
They are sometimes also referred to as digital SiPMs.
Unlike SiPMs, SPAD arrays integrate readout electronics directly
on-chip~\cite{Rochas_2003_First_fully_integrated}.
Consequently, a single data line allows the readout of millions of pixels, rather than requiring an independent analogue channel and digitisation chain for each.
They are well-suited for applications including
LIDAR~\cite{DBLP:journals/jssc/ZhangLAPWC19},
non-line-of-sight Imaging~\cite{spad-nonlineofsightimaging},
Raman spectroscopy~\cite{spad-biophotonics},
and event cameras~\cite{10655318}.
SPAD arrays featuring pixel pitches as small as
$2.5~\mu\text{m}$~\cite{Morimoto_OPEX:20,Shimada:22},
or equipped with time-to-digital converters (TDCs) providing
timestamp resolutions down to
$50~\text{ps}$~\cite{DBLP:journals/jssc/ZhangLAPWC19,Fischer:2022sfe},
have been demonstrated in the literature.
Recently, fast CMOS SPAD array sensors have been proposed as a
cost-effective light readout system for scintillating fibre-based
particle detectors~\cite{Franks:2023ttv}.

The next step in scintillator detectors involves 3D imaging of
photon-starved particle events in monolithic volumes.
Ref.~\cite{Bocchieri_2024_Scintillation_event_imaging} was able to
reconstruct the position of gamma rays with a SPAD array sensor.
However, the spatial resolution was limited by the use of the circle
of confusion method, applicable only to point-like sources.
Other independent works, based on Monte Carlo (MC) simulation
studies, have proposed more complex optical systems adopting
conventional photosensors to enhance the sensitivity to the spatial
depth of particle interactions, showing potential resolutions around
1 to 10 cm, depending on the volume size:
array of multiple lenses coupled to photomultiplier tubes (PMTs) for
a large-scale unsegmented liquid scintillator detector~\cite{Dalmasson:2017pow};
coded masks~\cite{infn-coded-mask}
or
objective lenses~\cite{CICERO2025170801}
coupled to SiPM to detect vacuum ultraviolet (VUV) scintillation
light in liquid argon.
In Ref.~\cite{Pisanti_2024}, a system of three objective lenses
viewing three orthogonal faces of a few-centimetre plastic
scintillator volume for fast neutron detection was studied with MC simulations.

The device with the greatest potential for 3D imaging of
photon-starved events is the plenoptic camera. It combines a single
main objective lens (\mainlens) with a \microlens array (\mla) placed
in front of an imaging photosensor. Each \microlens projected onto a
subset of sensor pixels acts as a tiny camera with a slightly
different perspective. This enables stereoscopic vision and thus
depth reconstruction from a single exposure.
A plenoptic camera is capable of capturing the ``light field'',
described by the function $\mathcal{L}(x,y,z,\phi,\theta)$, which
encodes the light intensity at each spatial point $(x,y,z)$
propagating in direction
$(\phi,\theta)$~\cite{Levoy_1996_Light_field_rendering}. For this
reason, it is often referred to as a light-field camera.
Although the concept dates back to
1908~\cite{Lippmann_1908_Epreuves_reversibles_donnant}, it is only
within the past two decades that the computationally intensive image
post-processing has become
practical~\cite{Adelson_1992_Single_lens_stereo,2005_Ng_Handheld_lightfield,2006_Ng_Digital_Lightfield_Photography}.
Alternative plenoptic configurations have been
proposed~\cite{Perwass_2012_Single_lens_3D_camera} and
commercialised~\cite{raytrix}.
Recently, a plenoptic camera instrumented with a charge-coupled
device (CCD) sensor has been proposed for 3D particle
dosimetry~\cite{Goulet_2014_Novel_full_3D}; however, it lacks the
timing information and single-photon sensitivity required to resolve
images of individual particles.
Closely related to the concept of plenoptic cameras,
Ref.~\cite{10.1117/12.2055951,Lebrun:2015gH} investigates a design
comprising lens arrays placed on the six faces of a 60~cm plastic
scintillator cube, read out using SPAD array imagers, studied through
MC simulations for gamma-ray detection via Compton-scattered
electrons. As noted by the authors, a principal limitation of the
approach is that tracks exceeding beyond the field of view of a
single lens cannot be accurately reconstructed, thereby restricting
the detector's capability to the tracking of particles below a
certain energy threshold.
A plenoptic system is mentioned as an alternative option, but was
neither conceptualised nor studied.
As of now, the application of plenoptic cameras with sub-nanosecond
timing to image particle events remains unexplored.

\begin{figure*}[htpb!]
  \centering
  \includegraphics[width=0.99\linewidth]{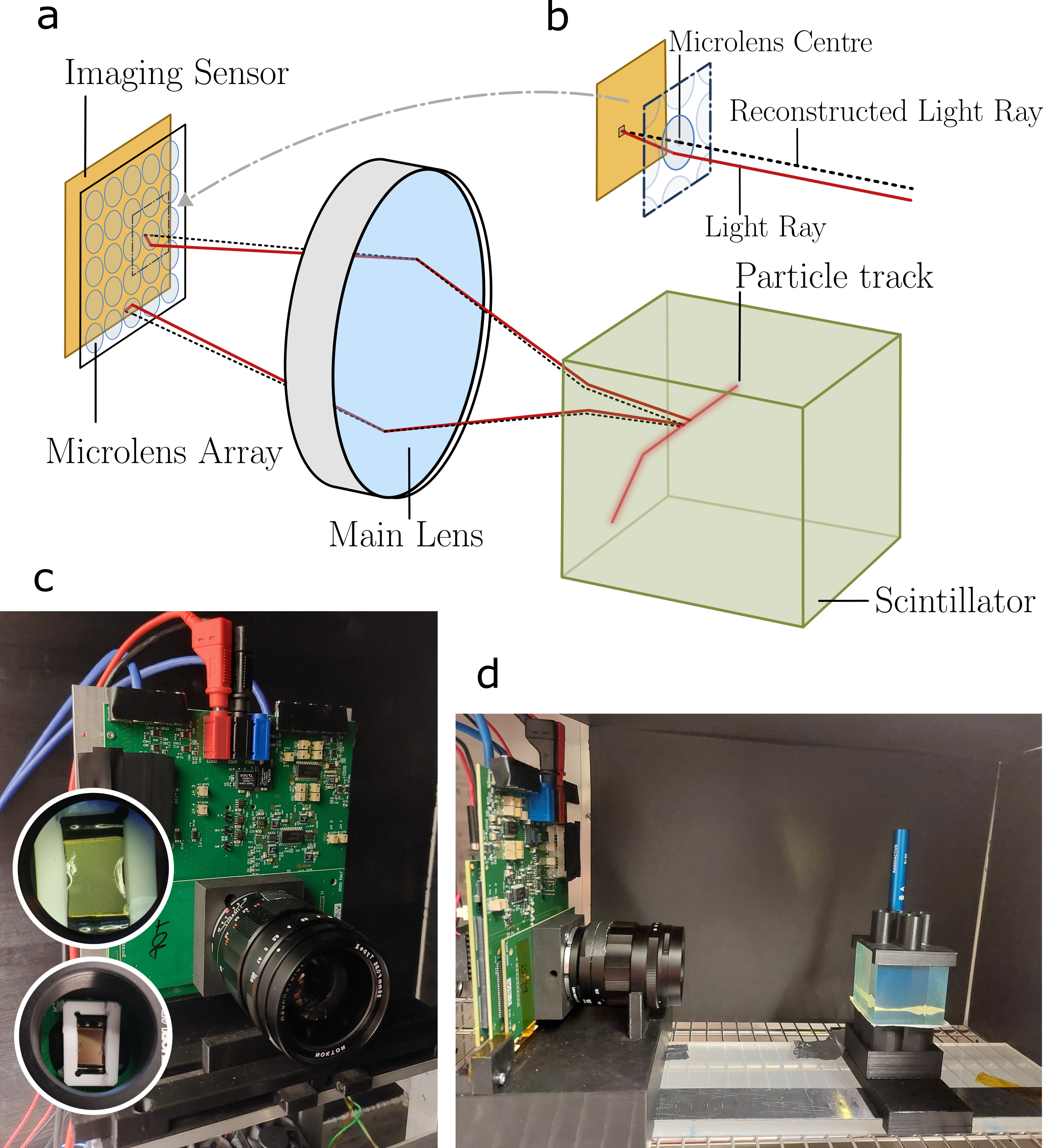}
  \vspace{0.2cm}
  \caption{\textbf{PLATON detection and reconstruction principle and
    the PLATON camera prototype.}
    \textbf{a} The PLATON imaging principle. A plenoptic camera
    detects photons emitted by a charged particle traversing the scintillator.
    \textbf{b} Principle used to reconstruct single photon
    micro-images, using the \microlens centre to determine the angle
    of the arriving photon.
    The intersection points of multiple photons are then used to
    reconstruct the particle track.
    \textbf{c} The PLATON camera prototype. Two insets show the
    ceramic \mla mount and a microscopy shot of the \mla itself.
    Here, the spill of the adhesive used to mount the MLA to the
    ceramic frame on the MLA can be seen on two sides.
    \textbf{d} The experimental setup, measuring \SrSource-electrons
    using the PLATON-prototype.
  }
  \label{fig:platon-prototype}
  \label{fig:pyramids}
\end{figure*}

In this article, we propose and demonstrate a paradigm shift in the
detection of particles in a large unsegmented scintillating volume,
resulting in high-resolution tracking and calorimetry of multi-particle events.
The detector consists of a system of multiple plenoptic cameras,
each one with an \mla behind the \mainlens and a SPAD array sensor
with time-resolving capabilities.
A pixel pitch of the order of 10~$\mu$m is desirable to constrain the
light field properly.
The post-processing of the 2D photon-starved images captured by the
plenoptic cameras provides a pure 3D image by ray-tracing each
detected photon back to its origin, leveraging the acquired light
field information.
If an organic scintillator with a decay time of a few nanoseconds is
employed, a single-pixel timestamp resolution on the order of a few
hundred picoseconds enables the effective suppression of the dominant
fraction of dark counts via time-coincidence techniques.
More generally, sub-nanosecond time resolution is often required in
particle physics experiments to obtain a high-purity signal sample.
The attenuation length, in both its scattering and
absorption/emission components, shall be much longer than the size of
the detector such that the original directions of the detected
photons are preserved.
This is easily achievable in organic scintillators with attenuation lengths of
a few metres for plastics~\cite{eljen-catalogue-2025}
and a few tens of metres for liquids~\cite{Djurcic:2015vqa,ABUSLEME2021164823}.
In this work, we demonstrate a detector concept with a plenoptic camera instrumented with a SPAD array, and
complement it with realistic MC simulation studies.
We developed different methods for image post-processing and a
data-driven calibration of the optical model.
The first prototype of this detector concept, which we named PLATON-prototype
(PLenoptic imAge of Tracked photONs), is shown in
Fig.~\ref{fig:platon-prototype}.
Based on the results of the prototype measurement campaign, we
provide the recipe of a future PLATON detector, whose development is
underway, determining the design parameters of both the plenoptic
system and the SPAD array sensor.
We
studied its design with a MC simulation of both a large cubic metre
size detector as well as in a smaller PLATON-10cm module exposed to an
accelerator neutrino beam.
A deep neural network (\neuralnetwork), based on the transformer
architecture developed initially for large-language
models~\cite{vaswani2017attention, 10278387},
was developed to precisely capture both the lateral and depth spatial
information with sub-millimetre resolution.
This work lays the foundation for an entirely different approach to
detecting elementary particles in dense materials, offering
exceptionally high spatial and temporal resolution.

\section{Results}
\label{sec:results}

\subsection{The PLATON prototype}
\label{sec:results-platon-prototype}

The PLATON-prototype, shown in Fig.~\ref{fig:platon-prototype},
adopts SwissSPAD2~\cite{Ulku_2019_A_512_x} as the imaging sensor.
It is a gated SPAD array with a 16.39 $\mu $m pixel pitch, a peak
photon-detection efficiency of 5\% at 520 nm (photon-detection
  probability of 50\% at 520~nm and a fill factor of 10.5\%, and no
on-chip per-pixel micro-lenses) and a median dark count rate of 0.26
counts per second (cps) per $\mu \text{m}^2$ of active area.
A single image frame corresponds to a gate window that can be varied
between 10 ns and 100 $\mu$s.
The SPAD array returns single-bit frames, where an active pixel
indicates that its SPAD has been activated, either by a photon or thermal noise.
2D intensity images with greyscale information are obtained from a
stack of multiple frames.
The plenoptic system was made in collaboration with Raytrix
GmbH~\cite{raytrix}, which designed and mounted the \mla (f/2.4,
\microlens diameter of 125 $\mu$m).
During the assembly process, some glue leaked onto the \mla, making
small areas on the left and right sides unusable for accurate imaging.
This can be seen in the inset of Fig.~\ref{fig:platon-prototype}.
A faster \mla, such as f/1, desirable to maximise light collection
efficiency, could not be used because the glob top, applied to
protect the wire bonds of SwissSPAD2, was too thick to place the \mla
sufficiently close to the sensor.
This limitation will be overcome in a future prototype.
A Voigtl\"ander Nokton II 25~mm photographic lens, refitted with a
C-Mount, was attached to the printed circuit board, on which the
sensor is situated.
The focus is set at a 30~cm distance to operate in a focused
plenoptic camera mode, with a virtual image being formed behind the sensor.
It was operated with f/2.4 to match the f-number of the \mla.
With these specifications, we expect a depth resolution of the order
of 1 to 5~mm and a sub-mm lateral resolution with a depth of field of
150~mm, calculated based on the effective resolution ratio (ERR) and
the overall minimal change in virtual depth, as described
in~\cite{Perwass_2012_Single_lens_3D_camera}.
In the future, a better resolution will be achieved with a higher
light collection efficiency and optics, currently constrained by the
slower \mla f-number.

\subsection{3D spatial resolution for a point-like light source}
\label{sec:results-spatial-resolution}

\begin{figure*}[htbp]
  \centering
  \includegraphics[width=1.0\linewidth]{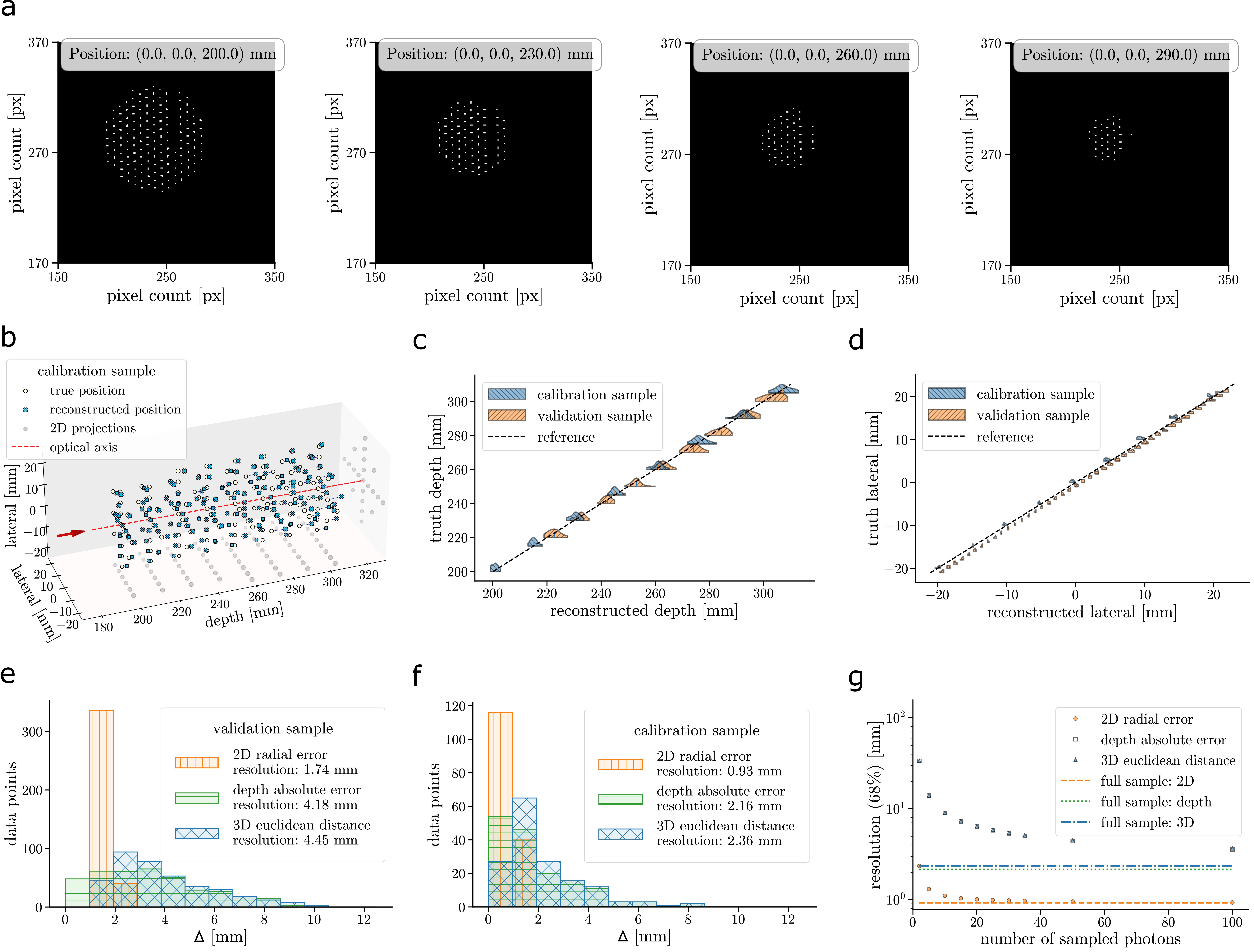}
  \caption{\textbf{Position reconstruction results for a point-like
    light source with the PLATON-prototype.}
    \textbf{a} Cropped images of the data taken during the
    calibration measurements. Visible is the calibration point source
    at different depths, on the optical axis; in classical imaging,
    the number of micro-images is used to infer the depth of the imaged objects.
    \textbf{b} True (white circles) and reconstructed (blue crosses)
    positions of the back-lit point light source with respect to the
    PLATON-prototype. Grey circles indicate the steps done by the
    moving stage projected onto the two perpendicular planes.
    The ``violin'' plots of the reconstructed depth (\textbf{c}) and
    lateral position (\textbf{d}) of the light source are shown. The
    dashed line corresponds to the case of a perfect reconstruction.
    The probability density function of the source depth or lateral
    position is shown for the calibration (blue) and validation
    (yellow) samples.
    The one-sided distributions of the residuals of the measured 3D
    positions obtained from the validation (\textbf{e}) and
    calibration (\textbf{f}) samples are shown.
    \textbf{g} The depth (green), 2D radial (orange), and 3D (blue)
    spatial resolutions as a function of the number of photons
    sampled from the calibration sample are shown on bottom right.
    The dashed horizontal lines correspond to the resolutions for
    high yield obtained from \textbf{f}.
  }
  \label{fig:calibration}
  \label{fig:reco-ptlike-source}
\end{figure*}

Using a back-illuminated pin-hole ($\diameter 50\,\mu\text{m}$)
attached to a motorised movement stage, two samples of data were collected.
One was used for calibration, while the other was used for validating
the post-processing method.
To achieve this, the pinhole was positioned at 168 discrete locations
relative to the camera using the motorised stage. These locations are
illustrated in Fig.~\ref{fig:calibration}. Since the primary
objective of this study was to calibrate the camera and evaluate the
achievable spatial resolution, these measurements were performed
without a scintillator, which would require a simple correction for
the refractive index.

The image post-processing was performed as described in
Sec.~\ref{sec:results-postprocessing}.
The calibration of the PLATON-prototype is discussed in
Sec.~\ref{sec:results-calibration}.
The results of the calibration and the spatial resolution obtained
with light intensity images, as well as a function of the number of
counts, are shown in Fig.~\ref{fig:calibration}.
The reconstructed 3D position of the light source
was compared to its nominal position on the stage.
Justified by its size, significantly smaller than the nominal lateral
resolution of the PLATON-prototype,
the source was assumed to be point-like.
The spatial residual of the depth, 2D and 3D position of the light
source is defined as the difference between the true (``t'') and the
reconstructed position (``r''):
\begin{align}
  & \Delta_{depth} = \sqrt{ \left(z_r - z_t\right)^2} \\
  & \Delta_{rad}   = \sqrt{ \left(x_r - x_t\right)^2 + \left(y_r -
  y_t\right)^2} \\
  & \Delta_{3D}    = \sqrt{ \left(x_r - x_t\right)^2 + \left(y_r -
  y_t\right)^2 + \left(z_r - z_t\right)^2}
\end{align}
where $z$ is the depth along the optical axis of the PLATON-prototype,
%\sout{camera},
while $x$ and $y$ are the lateral dimensions. The resolution is
defined as the 68\% 1-sided integral of the residual distribution.
The depth, 2D lateral and 3D resolutions in the regime of
fully-illuminated frames, i.e. when the light field is captured to
the best capability of the camera,
are, respectively, 4.2~mm, 1.8 mm and 4.5 mm for the validation
sample and 2.2 mm, 0.9 mm and 2.4 mm for the calibration sample.
The difference in resolution between the validation and calibration
data was found to be affected by remounting the camera on the
movement stage between the calibration and validation data runs.
This can be seen in the shift in the lateral reconstruction in
Fig.~\ref{fig:calibration}.
Thus, we will refer to the calibration sample to highlight the
intrinsic resolution of the PLATON-prototype.

The spatial resolution was also studied as a function of the number
of detected photons. A data-driven procedure was adopted: 1,000
photon-starved images were obtained by sampling the pixel position of
a certain predetermined number of counts, using the intensity image
of the point light source at a specific position as a PDF.
With 10 detected photons, the 2D lateral resolution falls below $1.2~\text{mm}$.
As expected, the 3D resolution is dominated by the depth one and
becomes worse than 10 mm for fewer than 10 photons.
The obtained spatial resolution is consistent with what is expected
from the design specifications of the PLATON-prototype, 
%\sout{camera}, 
including the
reduced FoV (see Sec.~\ref{sec:results-platon-prototype}), as well as
with results obtained with the optical simulation (see
Sec.~\ref{sec:optical-model}).
Using the simulated equivalent of the calibration sample, the
extracted depth resolution is 2.25~mm, with a 2D radial resolution of
0.5~mm. When accounting for possible systematic uncertainties
inherent in the stage and camera setup, these values are considered
consistent with those obtained from measurements.

An alternative reconstruction method for point-like sources was also tested using a similarly acquired data sample. This method is based on evaluating the likelihood of the measured light patterns with respect to a reference data sample.
A depth resolution of 1.5 mm was achieved along with a lateral
resolution of 0.1 mm.
With the sampled photon-starved images, the depth and lateral
resolutions decreased to, respectively, below 10~mm and 0.3~mm for 5 photons.
For fewer than 10 photons, depth ambiguities arise between points
located along a line extending from the principal point, as these
produce nearly concentric image patterns.
For further details, see Sec.~\ref{sec:likelihood-analysis}.
It is worth noting that the likelihood study should not be
interpreted as the achieved spatial resolution, as it would be
impractical to build a PDF for every single possible point down to
the relevant precision in a relatively large volume.
On the other hand, the largely improved resolution compared to the
reference method described in Sec.~\ref{sec:results-spatial-resolution}
(better than a factor of 2 or 3 for more than 30 counts)
indicates that the 2D frames provided by the PLATON-prototype contain
more spatial information than what can be exploited with the
developed post-processing method.
Such a feature highlights that a more advanced post-processing, such
as one that uses prior information of the optical model more detailed
than that described in Sec.~\ref{sec:results-postprocessing} or deep
learning (see Sec.~\ref{sec:results-platon-nu-neutrino} and
\ref{sec:methods-nn}) would further improve the 3D spatial resolution.
Our near-future plan is to pursue both directions further.

\subsubsection{Detection of \SrSource electron events}
\label{sec:results-90Sr-reco-point}

\begin{figure*}[htpb]
  \centering
  \includegraphics[width=0.95\linewidth]{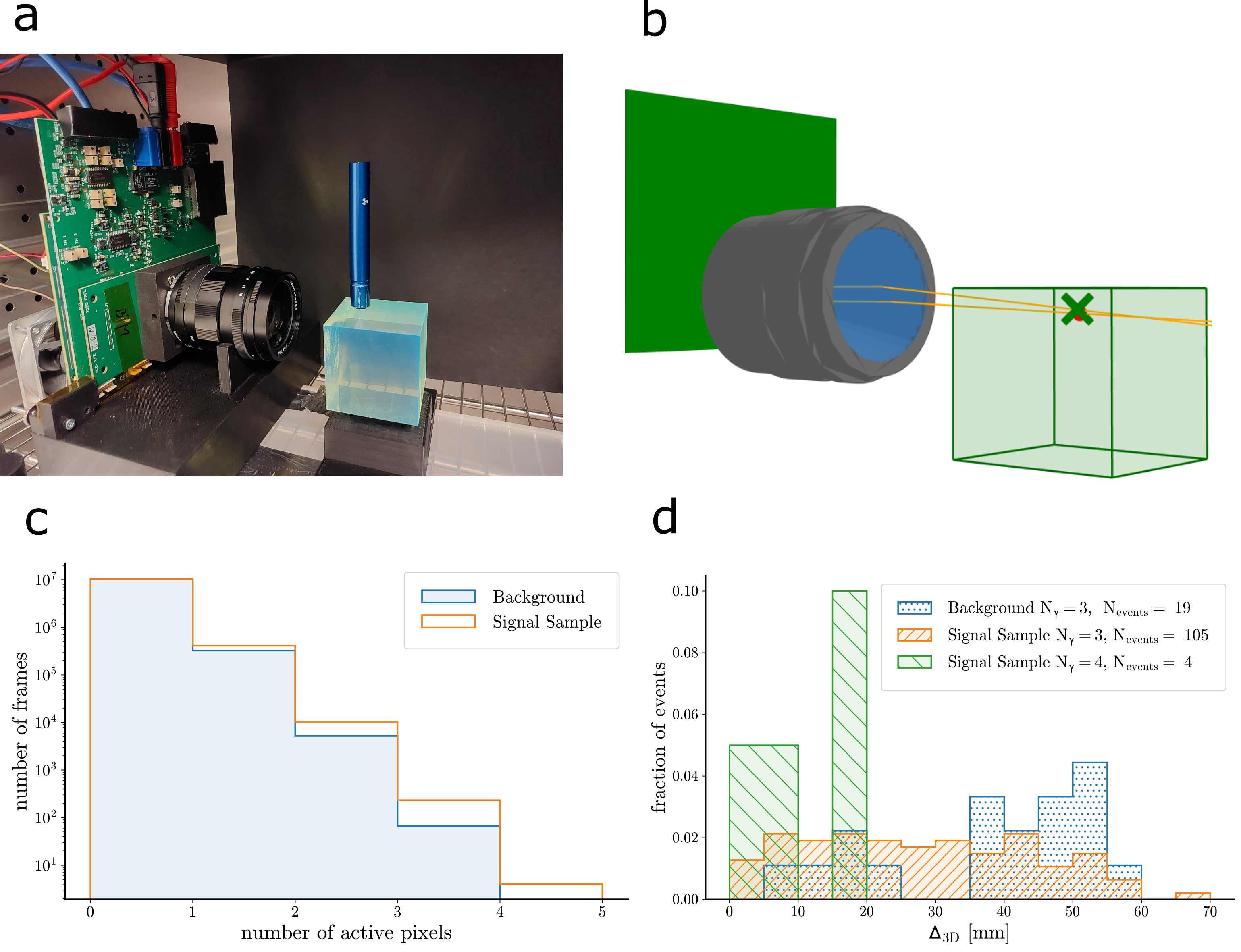}
  \caption{\textbf{Detection of $\beta$ electrons from a \SrSource
    using the PLATON-prototype.}
    \textbf{a} PLATON-prototype
    pointing to the scintillator
    block exposed to the \SrSource positioned on the top face.
    \textbf{b} A candidate electron event
    in the scintillator (green volume) selected from the \SrSource
    sample. The prototype \mainlens, the reconstructed ray trace of
    the scintillation photons (orange lines), the nominal position of
    the source (red dot) and the reconstructed position of the
    electron (green cross) are shown.
    Other $^{90}$Sr events can be found in Fig.~\ref{fig:90Sr-events}.
    \textbf{c} Distribution of the number of counts per frame for the
    \SrSource (orange) and the background (blue) samples.
    \textbf{d} Residual of the reconstructed \SrSource electron
    position from the \SrSource sample with 4 counts (green), 3
    counts (orange) and from background sample (blue) with 3 counts.
  }
  \label{fig:sr90}
\end{figure*}

The final measurement with the PLATON-prototype aimed to detect, with a plenoptic camera, single elementary
particles in plastic scintillator on an event-by-event basis.
A $50 \text{ mm } \times 50 \text{ mm } \times 50 \text{ mm }$ block
of EJ-262 plastic scintillator~\cite{datasheet_EJ_260_262}
(parameters are listed in Sec.~\ref{sec:optical-model}) was exposed
to a \SrSource source, placed on the top side. The electrons exiting
the container of the source can carry an energy up to a maximum of 1.5 MeV.
In order to maximise the photon collection, we had to position the
scintillator block about 5 cm from the \mainlens of the PLATON-prototype, just outside the optimal depth of field (150 mm), and hence out of focus. This is limiting the angular resolution on the photon direction.
To certify the reconstruction of the electron position, two different
datasets were collected, with and without the \SrSource on top of the
scintillator block. We name the two datasets, respectively, as
\SrSource and background.
To decrease the probability of dark counts, the PLATON-prototype was
placed in a thermal chamber at a temperature of $-5^{\circ}$C,
reducing the median dark count rate (DCR) to 0.002 cps/$\mu$m$^2$
(0.4 cps/pixel).
The background sample allows us to identify noisy pixels, i.e. with a
DCR higher than
0.5 cps/pixel,
which were masked at the post-processing stage, further reducing the SwissSPAD2
photodetection efficiency (PDE)
down to 2.5\%, however increasing the signal-to-noise ratio.
Thus, given the limited light yield of the prototype (f/2.4, active
area reduced by the glue), only frames with at least four counts were
selected, as these are more likely to contain at least two
scintillation photons, necessary to fit a point in the object space.
In the post-processing (see Sec.~\ref{sec:results-postprocessing}),
candidate photons were identified if the relative minimum distance
was less than 3~mm.
Out of 9 frames with four counts found in the \SrSource sample, only
four events were selected upon the convergence of the post-processing.
As shown in Fig.~\ref{fig:sr90}d, for each event the reconstructed
position of \SrSource was closer than 20 mm to its true position,
dominated by the depth resolution, as discussed in
Sec.~\ref{sec:results-spatial-resolution}. 

This result is consistent, within statistical uncertainty, with the spatial resolution obtained for a low number of photons when accounting for the distance between the main lens and the scintillator, which is not in focus, as shown in Fig.~\ref{fig:reco-ptlike-source}.

\begin{figure*}[htbp]
  \centering
  \includegraphics[width=0.9\linewidth]{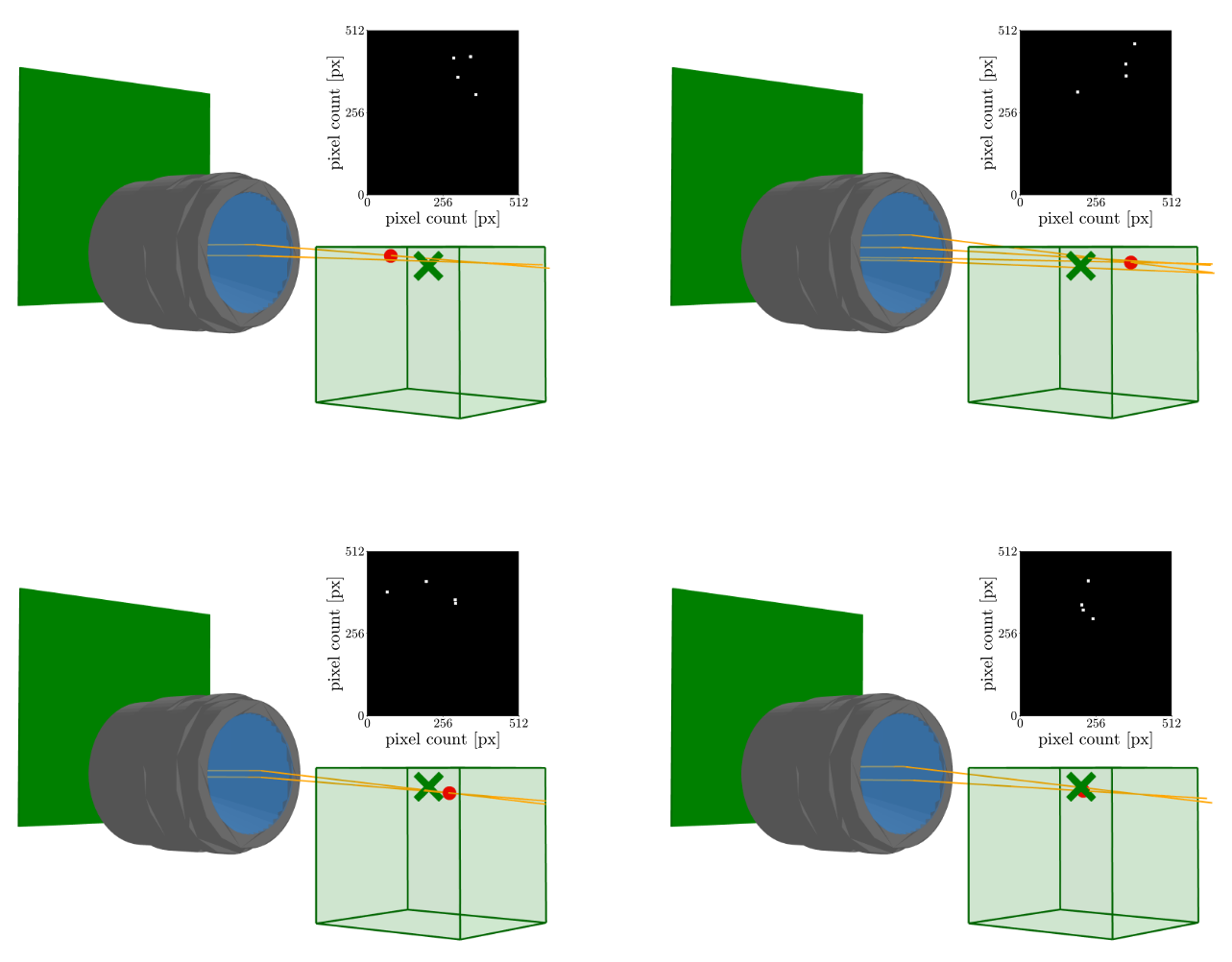}
  \caption{\textbf{Four event displays of reconstructed \SrSource
    electron candidates.}
    Shown are the four identified events, with 4 non masked initial
    counts on the sensor.
    For each event the image on the sensor is show, with enlarged
    active pixels for visibility.
    Below each event is shown in an event display, where in orange,
    the reconstructed rays are visible, together with the real source
    location (green cross) and the reconstructed point of the
    \SrSource (red circle). The data analysis is described in
    Sec.~\ref{sec:results-90Sr-reco-point}.
  }
  \label{fig:90Sr-events}
\end{figure*}

The same analysis was done on both the \SrSource and the background
sample with frames containing 3 counts.
The rejection rates were, respectively, 61\% and 81\%.
In the latter two cases, the distribution of the distance from the
\SrSource nominal position is evenly distributed across the entire depth range.
Event displays of the four \SrSource candidates can be found in
Fig.~\ref{fig:90Sr-events}.
From a toy Monte Carlo simulation of the background sample we expect 0.17 background events with 4 or more counts to be reconstructed between 0 and 20 mm from the true \SrSource position.

\subsection{Simulated particle detection with the PLATON-10cm detector}
\label{sec:results-platon-nu}

\begin{figure*}[htbp]
  \centering
  \includegraphics[width=0.95\linewidth]{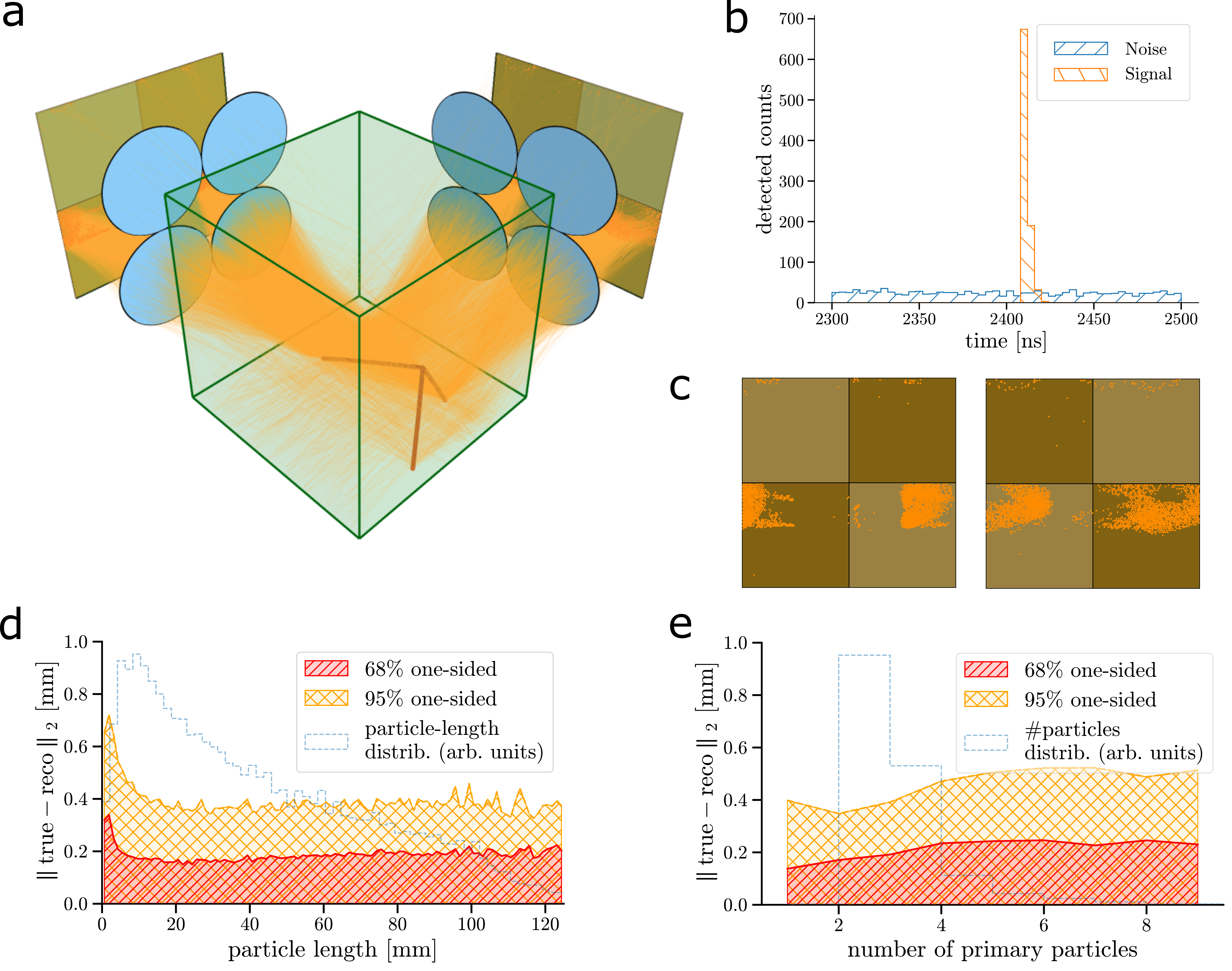}
  \caption{\textbf{Simulation and evaluation results of the proposed
    PLATON-10cm detector module.}
    \textbf{a} Simulated interaction of a muon neutrino in the PLATON-10cm
    module. Both the truth particle tracks and the ray traces of the
    detected photons along with their arrival positions at the SPAD
    array are shown in different tones of orange. The count time
    distributions for signal photons (orange) and dark counts (blue)
    are shown in \textbf{b}, while \textbf{c} illustrates the frames
    collected by the SPAD array sensors.
    3D spatial resolution as a function of the particle range
    (\textbf{d}) and the number of particles in the event
    (\textbf{e}). Respectively, the distribution of the particle
    range (\textbf{d}) and of the number of particles per event
    (\textbf{e}) is shown (dashed light blue).
  }
  \label{fig:neutrino-sim}
  \label{fig:neutrino-postprocessing}
  \label{fig:neutrino-nn}
\end{figure*}

As a physics study case, we chose the detection of GeV muon neutrinos.
In fact, the PLATON concept could be suitable for the near detector of future
accelerator long-baseline neutrino oscillation (LBL) experiments,
which will search for leptonic charge-parity violation and determine
the neutrino mass ordering
~\cite{Hyper-Kamiokande:2025fci,Hyper-Kamiokande:2018ofw, Abi:2020qib}.
In these experiments, the precise measurement of the neutrino-nucleus
cross section is crucial for the reduction of the key systematic
uncertainties, and it relies on the high-resolution tracking and
calorimetry of final-state low-momentum hadrons (e.g., multiple
protons down to 200 MeV/c).
We studied the performance of a small PLATON module consisting of two
optical arrays, each of four plenoptic cameras, pointing to two
orthogonal faces of a $10 \times 10 \times 10~\text{cm}^3$ block of
EJ-262 plastic scintillator (parameters are listed in
Sec.~\ref{sec:optical-model}), which we call PLATON-10cm.
Performing the analysis described below would be challenging for a
larger detector volume with the computing power currently available.
Since the attenuation length is way longer than the scintillator
volume, a perfect transparency was assumed.
This assumption is justified even for larger volumes due to the
availability of organic liquid scintillators with analogous light
output but attenuation lengths
of the order of 20 m or greater~\cite{JUNO:2021vlw,ABUSLEME2021164823}.
The simulated detector is shown in Fig.~\ref{fig:neutrino-sim}a.
Owing to the results obtained with the PLATON-10cm prototype (see
Sec.~\ref{sec:results-platon-prototype}), we designed the PLATON-10cm
detector unit (a single camera).
The adopted strategy
consisted of optimising the lateral resolution, with a depth
resolution of 1~mm over a depth of field of 10~cm.
Exploiting the depth resolution to match the two orthogonal views,
the result is a 3D spatial resolution that mainly belongs to the
sub-millimetre lateral one, especially in relatively low-multiplicity
events such as neutrino interactions.
Both the \mla and the \mainlens were simulated with f/1. The diameter
of the \mainlens is 50 mm, while that of the \microlens{es} is 1 mm.
Realistic parameters of an optimised SPAD array sensor, which we
refer to as PLATON SPAD, were simulated.
The PDP (40\% peak at 500 nm), simulated as in
Sec.~\ref{sec:optical-model} and the pixel pitch of $25~\mu\text{m}$
with a 60\% geometrical fill factor amount to a PDE of 24\%. We
neglect the tiling between adjacent photosensors.
With all efficiencies accounted for, 3,670 photons were detected per event.
The SPAD time structure shown in Fig.~\ref{fig:neutrino-sim}b was simulated.
Each count is measured with a 200~ps resolution timestamp.
R\&D on PLATON SPAD is ongoing to achieve the parameters
above~\cite{Kaneyasu2025PlatonSPAD}.

\subsubsection{Single-point spatial resolution of an array of multiple cameras}
\label{sec:results-platon-nu-resolution-point}

Before studying the capability of a PLATON-10cm module to detect
neutrinos, the 3D spatial resolution for a point light source
equivalent to 1 MeV energy deposition was evaluated.
Of the 10,000 photons produced, on average 141 photons were detected.
The 3D spatial resolution was found to be 0.3 mm.
For completeness, we compared plenoptic cameras with classical
cameras, obtained by removing the \mla and adapting the lens-sensor
distance, focusing the camera on the centre of the scintillator.
Here an average of 149 photons were detected per event.
While a single classical camera cannot precisely reconstruct the 3D
position of the light source, we found that the resolution provided
by multiple classical cameras is about four times worse than that of
plenoptic cameras.
Comparing a setup with two plenoptic cameras to a setup with two classical cameras, we see a significant difference in reconstruction efficiency, indicated in Fig.~\ref{fig:CC_PC}.
While the two plenoptic cameras are able in all events to reconstruct the 3D origin of the point source, when the event only falls in the view of a single classical camera, the reconstruction fails, leading to around 23\% of failed reconstruction attempts.
On the other hand, we note that the size of the scintillator block is
not that large and does not require a very long depth of field (DoF),
making the performance of an array of classical cameras not too far
from that of plenoptic cameras. As we will see later in
Sec.~\ref{sec:results-platon-nu-resolution-point-large}, the
resolution of a system of classical cameras deteriorates compared to
plenoptic cameras for a m$^3$ volume.
Thus, we focus on the PLATON-10cm configuration with plenoptic cameras, as
described in the previous section.
For more details, we refer to
Sec.~\ref{sec:results-platon-nu-resolution-point-details-and-classical}.

\subsubsection{Neutrino detection in PLATON-10cm}
\label{sec:results-platon-nu-neutrino}

The neutrino flux of the T2K LBL
experiment~\cite{Abe_2015_Measurements_of_neutrino,
FluxPrediction2016} was simulated parallel to both sensor planes, and
uniformly in the scintillator volume with the NEUT 5.5.0 event
generator~\cite{Hayato_2021_The_NEUT_neutrino}.
Assuming that an external detector, such as a time projection chamber with a spatial resolution and angular resolution of 1~mm and 0.05~rad respectively
in a magnetised volume,
would be able to detect and identify the muon escaping the
scintillating volume,
we simulated a sample of charged-current (CC)
neutrino interactions, disregarding the contribution of
neutral-current (NC) events.

\begin{figure*}[htbp]
  \centering
  \includegraphics[width=1.0\linewidth]{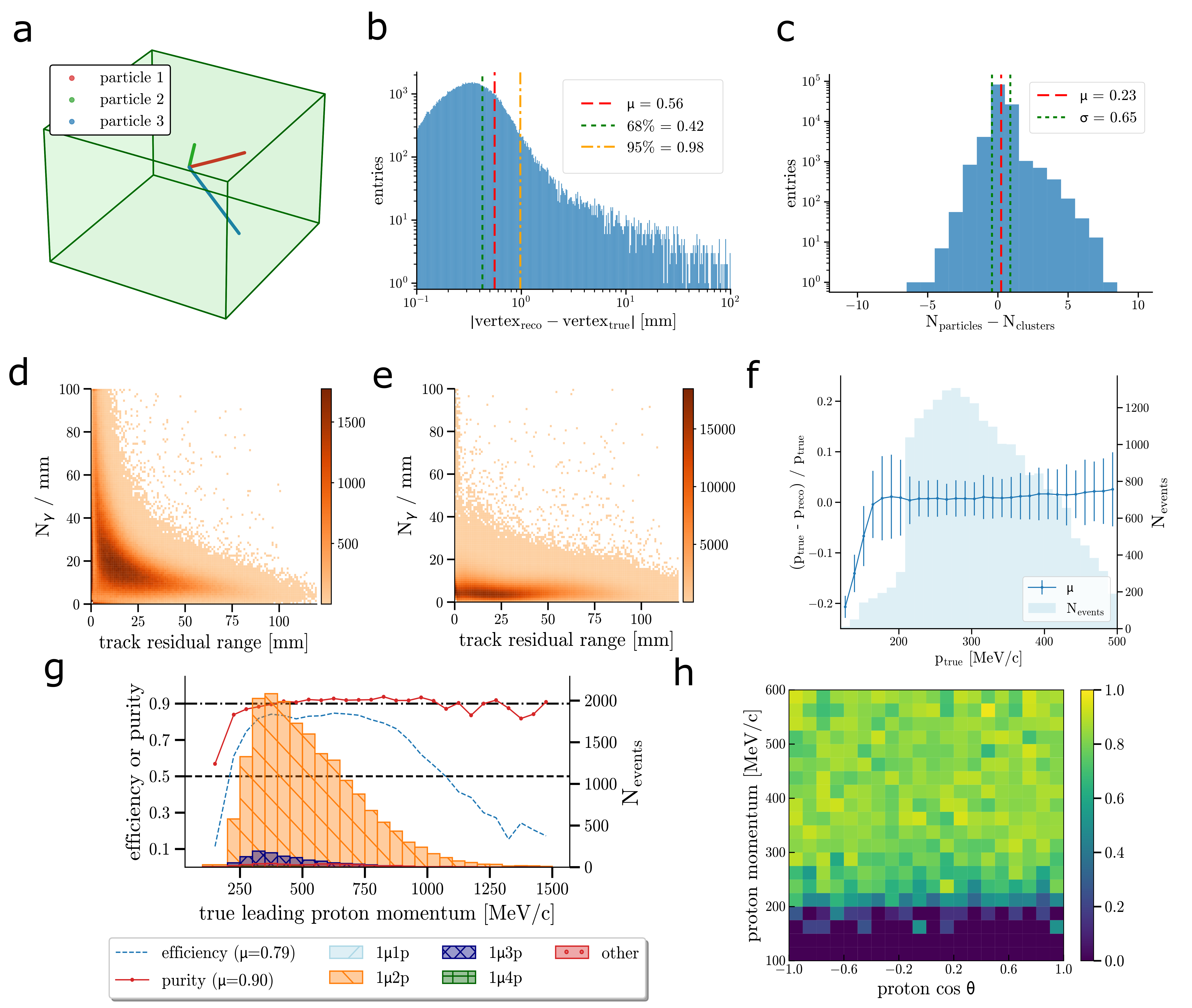}
  \caption{
    \textbf{Simulated muon neutrino event analysis results.}
    \textbf{a} Particle tracks reconstructed from a
    \ccOneMuZeroPiTwoProt neutrino interaction (a muon and two
    protons with no pions) after applying the image post-processing
    and the pattern recognition.
    \textbf{b} Residual of the reconstructed neutrino vertex position
    for proton events.
    \textbf{c} Difference between the number of particles produced by
    the neutrino interaction and the number of reconstructed clusters
    for muon events.
    The measured energy loss by a proton \textbf{d} and a muon
    \textbf{e} along the corresponding reconstructed track for all the events.
    \textbf{f} Proton momentum resolution (mean and 68\% std) as a
    function of the stopping proton true momentum for selected
    \ccOneMuZeroPiTwoProt events (dark blue line).
    The true momentum distribution (light blue) of the protons from
    all the neutrino events is also shown.
    \textbf{g} True momentum distribution of the leading proton in
    the \ccOneMuZeroPiTwoProt selected sample, including efficiency and purity.
    \textbf{h} Proton reconstruction efficiency (colour map) as a
    function of the stopping proton true momentum and angle after
    applying the \ccOneMuZeroPiTwoProt selection.
  }
  \label{fig:neutrino-pattern-recognition}
  \label{fig:neutrino-selection}
\end{figure*}

We developed a custom neural network based on the transformer
architecture, commonly used in generative large language models, to
efficiently capture the hyper-dimensional correlation between the
detected scintillation photons. Details on the neural network are
given in Sec.~\ref{sec:methods-nn}.
We evaluated the particle tracking performance by computing the
distance between each point (Geant4 output, see
Sec.~\ref{sec:optical-model}) along the truth trajectory of the
particle and the closest reconstructed point
(Figs.~\ref{fig:neutrino-nn}d and~\ref{fig:neutrino-nn}e).
Overall, an average 3D-tracking resolution of approximately
$190\,\mu\text{m}$ is obtained, with a range of $150\,\mu\text{m}$ to
$340\,\mu\text{m}$. When considering the distance between each
reconstructed point and the closest truth point, the resolution
becomes more consistent, with a narrower range of $160\,\mu\text{m}$
to $230\,\mu\text{m}$, though the average remains unchanged.
The resolution is smaller than $200\,\mu$m for events with three
particles or fewer.
When particle ranges are less than 3 mm, the resolution averages $180\,\mu$m.
Overall, the results demonstrate that a tracking accuracy well below
1~mm is achievable within a sizeable PLATON module.
This is in agreement with the results of~\ref{sec:likelihood-analysis} and~\ref{sec:results-platon-nu-resolution-point}.
A pattern recognition algorithm was developed to group the
\photorig{s} into different clusters that are then assigned to
different particles.
The results of the pattern recognition are shown
in Fig.~\ref{fig:neutrino-pattern-recognition}.
The interaction vertex resolution for the full sample of CC
interactions is 0.42 mm.
The average difference between the number of identified and true
particles is 0.25 with a standard deviation of 0.74. More details
about the implementation of pattern recognition and particle identification
can be found in
Sec.~\ref{sec:results-platon-nu-neutrino-details-pattern-recognition}.

\begin{figure*}[htpb]
  \centering
  \includegraphics[width=0.95\textwidth]{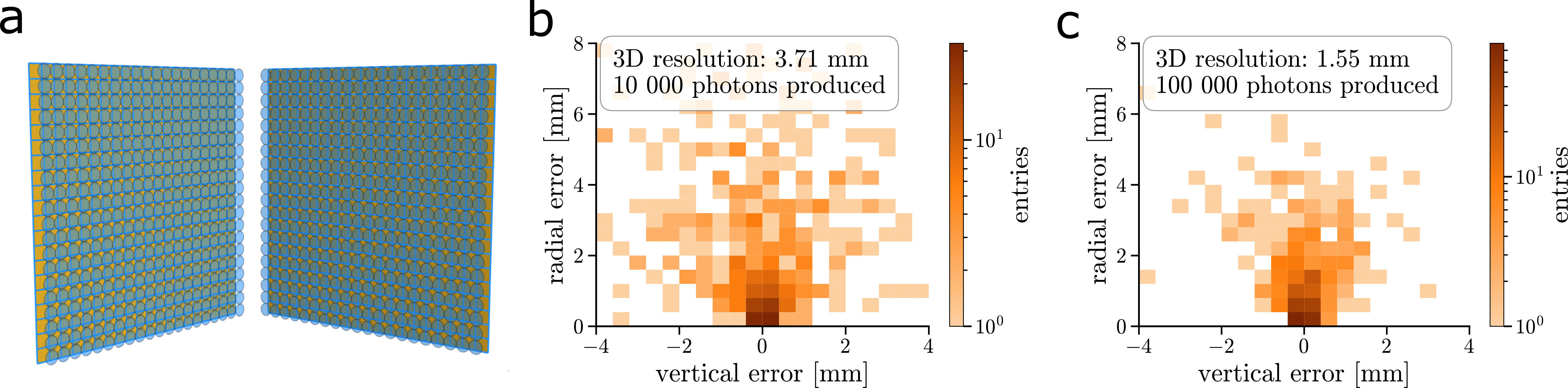}
  \caption{\textbf{Setup and reconstruction of a tonne-scale version
    using the PLATON detector concept.}
    \textbf{a} The geometry of a tonne-scale $1 \times 1 \times
    1~\text{m}^3$ PLATON-1m detector. The plenoptic cameras are shown in
    blue. The same optics configuration as in
    Sec.~\ref{sec:results-platon-nu-resolution-point} is adopted, resulting in a total of 800 plenoptic cameras, each with about
    $4\times10^{6}$ pixels ($\sim 3.2\times10^{9}$ pixels in total).
    The spatial resolution in cylindrical coordinates for a point
    light source emitting \textbf{b} 10,000 photons and \textbf{c}
    100,000 photons is shown.
  }
  \label{fig:neutrino-big-detector}
\end{figure*}

Two samples of candidate neutrino CC interactions were selected,
featuring either a muon and a proton with no pions in the final state
(\ccOneMuZeroPiOneProt), or a muon and two protons with no pions
(\ccOneMuZeroPiTwoProt).
The first one is the most common final-state topology below 1 GeV,
thus the dominant channel at the T2K and Hyper-Kamiokande LBL
experiments, and is mainly affected by CC quasi-elastic (CCQE)
interactions, accounting for approximately 37\% of all events.
The second sample is a golden channel to characterise nuclear
processes, such as 2 particle - 2 holes (2p2h) events or final state
interactions (FSI)
that can introduce biases in the reconstruction of the neutrino energy.
The proper theoretical modelling of these processes is a source of significant
uncertainties in LBL experiments. Thus, the precise detection of
\ccOneMuZeroPiTwoProt events is crucial for the precise measurement
of neutrino oscillations.
More information can be found in
Sec.~\ref{sec:results-platon-nu-neutrino-details-selection}, together
with the details of the neutrino event selection.

To demonstrate the capability of a PLATON-10cm detector module to detect
neutrinos, we focus on the selection of both \ccOneMuZeroPiOneProt
and \ccOneMuZeroPiTwoProt events, noting that the latter are
notoriously more challenging to identify due to the presence of
higher multiplicity.
From Fig.~\ref{fig:neutrino-selection}f, one can see that the
momentum of stopping protons reconstructed from the range is inferred
with a resolution better than 10\% across almost the entire range,
approximately 5\% at the energy flux peak, and close to 10\% only
below 200 MeV/c.
The \ccOneMuZeroPiTwoProt selection purity is 90\% up to 1.4 GeV/c
momentum of the most energetic (leading) proton. The small background
contamination comes mostly from events with higher proton
multiplicity, typically produced with lower momenta.
The total selection efficiency is 78\%. It exceeds 80\% for leading
proton momenta above 300~MeV/$c$, but drops below 50\% for momenta
above 1.1~GeV/$c$, where the muon and the proton become more collinear.
The proton reconstruction efficiency, shown for the
\ccOneMuZeroPiTwoProt sample, exhibits a 50\% threshold at
approximately 215 MeV/c.
For the case of  \ccOneMuZeroPiOneProt events,
both the selection efficiency and the purity reach 92\%.
The direction of single protons is determined with an angular
resolution of $1.5^{\circ}$.

The same analysis has also been conducted for a system of classical
cameras, as reported in
Sec.~\ref{sec:results-platon-nu-neutrino-details-selection-classical}.
When compared to plenoptic cameras, similar conclusions can be drawn,
as discussed in Secs.~\ref{sec:results-platon-nu-resolution-point}
and~\ref{sec:results-platon-nu-resolution-point-details-and-classical}.

\subsection{Single-point spatial resolution at 1 m$^3$ scale}
\label{sec:results-platon-nu-resolution-point-large}

Finally, the spatial resolution for a point light source in a PLATON
detector with a $1 \times 1 \times 1~\text{m}^3$ organic scintillator
has been studied, named PLATON-1m.
Due to limited computational resources, we were unable to perform a
neutrino simulation study similar to the one in
Sec.~\ref{sec:results-platon-nu-neutrino}, but we leave it for future work.
The same analysis as in
Sec.~\ref{sec:results-platon-nu-resolution-point} was performed.
As shown in Fig.~\ref{fig:neutrino-big-detector}, a 3D spatial
resolution of 3.7~mm can be achieved for an energy loss of $\sim 1$
MeV (10,000 photons generated). For reference, this resolution
corresponds to an effective segmentation of approximately 7.5~mm.
Unlike a segmented detector, it improves up to 1.5~mm for an energy
loss of $\sim 10$ MeV (100,000 photons generated), with an effective
segmentation of 3~mm.

The same analysis was performed with arrays of classical cameras. As
anticipated above, this time we found that for a 1 m$^3$ size
detector, the plenoptic system has a spatial resolution almost 4
times better than that of classical cameras.

With a few improvements, a sub-millimetre spatial resolution in a
$1~\text{m}^3$ scintillator could be achieved.
First, the optical parameters have not been optimised for such a
volume but have been kept the same as for the $10 \times 10 \times
10~\text{cm}^3$ module.
Moreover, the used post-processing method adopts a simplified
parametrisation of the \mla. On the other hand, we also observed that
the NN described in Sec.~\ref{sec:methods-nn} allowed for an
additional improvement in spatial resolution.
Thus, one might expect a similar improvement also in the tonne-scale PLATON-1m.
It is interesting to note that a smaller pixel pitch, though
technically challenging, would allow for better constraint of the
light field. Preliminary studies showed 30\% improvement in spatial
resolution for a 1 $\mu$m pitch without any adaptation of the optics.

\section{Discussion}
\label{sec:discussion-conclusions}

In this work, we propose a change of paradigm in the high-resolution
detection of elementary particles in dense scintillator-based detectors.
The key is the use of imaging photosensors with single-photon
sub-nanosecond time resolution,
such as SPAD array sensors, integrated into a system of plenoptic
cameras to provide unambiguous 3D images of elementary particles
interacting within a monolithic volume of scintillator. We referred
to this detector concept as PLATON.

We built and successfully characterised the spatial resolution of the
PLATON detector prototype, a plenoptic camera instrumented with a SPAD array sensor,
and we developed a post-processing method suitable for
photon-starved images.
We also succeeded in reconstructing the position of $^{90}$Sr
electrons on an event-by-event basis.
All the results are in agreement with our simulations, which gives us
confidence in designing a realistic PLATON detector for neutrino detection.

Building on the experience gained through prototyping, we carried out
a simulation-based analysis of muon neutrino interactions in a PLATON
detector, employing an improved yet realistic design.
We developed a high-performance image post-processing based on
deep learning and complemented by pattern recognition algorithms.
A tracking resolution down to 200~$\mu$m, along with high purities
and efficiencies in the selection of neutrino interactions with
multi-proton final states, was achieved.

The extrapolation to a 1-tonne scintillator detector showed a spatial
resolution of a few millimetres for point light sources. Although it
is already competitive with state-of-the-art plastic scintillator
detectors, we expect more sophisticated image post-processing, such
as deep learning,
or a plenoptic system optimised for a metre-scale volume,
to be promising solutions for achieving sub-millimetre spatial
resolution. Future developments, such as optics optimised for
metre-scale volume and smaller pixels will further improve the
spatial resolution.
This will be the topic of future dedicated work.

We compared the simulation results with those obtained using a system
of classical cameras and found that plenoptic cameras outperform
classical systems, regardless of the scintillator volume size.

The PLATON detector configuration offers broader applicability beyond
accelerator-based neutrino studies, which were adopted here as a case study.
Its calorimetric capability combined with high spatial resolution could make this concept potentially suitable for scintillator-based
neutrino-less double-beta decay (\nuless) experiments and makes it worth studying. 
Examples could be Ref.~\cite{SuperNEMO} or 
large-scale detectors such as Ref.~\cite{KamLAND-Zen:2024eml,SNO:2021xpa,AUTY2023168204,SNO:2020fhu} (see also Ref.~\cite{Dolinski:2019nrj} for a
comprehensive review).
Furthermore, following the design philosophy outlined in
Ref.~\cite{Dalmasson:2017pow}, the PLATON concept appears promising
for application in very-large-scale neutrino
detectors~\cite{DAYABAY,JUNO:2021vlw,SNO:2025qqy,Theia:2019non}.
Similarly, it may enhance the reconstruction of light rings
characteristic of Cherenkov radiation~\cite{Jiang:2019xwn,PEREGO2012606}.
Compared to liquid argon time projection chambers (LArTPCs), currently the only technology capable of 3D tracking particles produced by neutrino interactions with O(mm) resolution in giant volumes,  the PLATON technology would profit of its much faster response. Thus, it is ideal for the detection and the reconstruction of the time-of-flight of fast neutrons produced by neutrinos. %as well as for deployment in high rate environment. 
All would be achieved without the need of big cryogenic infrastructures.
On the other hand, its scalability to the kilotonne scale achieved by LArTPCs will have to be proven.

To conclude, our work paves the way towards high-resolution particle
tracking and calorimetry in unsegmented volumes.
The development of SPAD array imaging sensors, or other photosensor
technologies with analogous features, is the key.
We are currently conducting R\&D to develop a SPAD array sensor that
meets the requirements discussed in this article, specifically a PDE
similar to that of high-dynamic range SiPMs
~\cite{hamamatsu:mppc}
and single-photon sub-nanosecond time resolution.

Owing to high-spatial resolution in large and dense active volumes combined with an excellent time resolution, the PLATON concept therefore naturally opens up new areas of application for deploying plenoptic cameras for imaging, including positron-emission tomography (PET), neutron radiography, muon tomography, synthetic computed tomography (CT), and proton CT.

\section{Methods}
\label{sec:methods}

\subsection{Optical model and simulation}
\label{sec:optical-model}

The interaction and propagation of particles in the scintillator
volume were simulated with the Geant4 toolkit
~\cite{Allison_2006_Geant4_developments_and,Agostinelli_2003_Geant4_a_simulation_toolkit,Allison_2016_Recent_developments_in}.
The parameters of polyvinyltoluene-based plastic scintillator are simulated with
light yield (8,700 photons / MeV),  emission and absorption spectra,
attenuation length (2.5 m), refractive index (1.58), and decay time
($\sim 2.1$ ns) characteristic of EJ-262~\cite{datasheet_EJ_260_262},
the scintillator used for the PLATON-prototype.
It is worth mentioning that the purpose of this simulation study is general and is applicable to a wide variety of organic scintillators, including those in the form of liquid, often used in giant neutrino experiments \cite{JUNO:2021vlw, DAYABAY}.
Their typical light yield is around 10,000 photons / MeV, attenuation lengths equal or above 20 m, as well as very high purity, thus, a very uniform optical performance.

Neutrinos are simulated as follows. The neutrino energy was simulated according to the flux made public by the T2K experiment, where it is used in the neutrino oscillation data analyses~\cite{Abe_2015_Measurements_of_neutrino,
FluxPrediction2016}. The flux is peaked at 0.6 GeV and its broad band ranges mainly from 0.2 to 1.5 GeV, with a long tail up to 30 GeV. 
The T2K neutrino angular distribution is wide, with a standard deviation of about 4 meters at the near detector location \cite{T2K:2012bge}. Thus, for simplicity we decided to simulate neutrinos on a fixed direction, parallel to both sensor planes, without loss of generality.
The NEUT 5.5.0 neutrino event generator \cite{Hayato_2021_The_NEUT_neutrino}, used at the T2K experiment for neutrino data analyses,
is used to simulate the neutrino interactions in the scintillator. It provides the list of final-state particles (particle types and corresponding momentum vectors) and the 3D position of the neutrino interaction vertex.
Then, the propagation of the final-state particles in matter is simulated using the Geant4 toolkit
~\cite{Allison_2006_Geant4_developments_and,Agostinelli_2003_Geant4_a_simulation_toolkit,Allison_2016_Recent_developments_in}.
We assume the distance between the position of the ionising particle
and the final emission of the scintillation light to be small
compared to the 3D spatial resolution of the PLATON
camera~\cite{PDG,Kolanoski:2020ksk} and, thus, no additional smearing
on the position of the scintillation light emission is applied.
The optical photons produced by the scintillation process are propagated using native Geant4 optical photon processes, including absorption and reflections following the Fresnel equations. 
The speed of photons in scintillator is about 20 cm / ns. 
Once the photons leave the scintillator volume, their positions and directions are stored for subsequent processing steps.

The second part of the optical simulation is based on a custom
library that simulates lenses, sensors, and refractive surfaces
outside the scintillator volume. It performs the ray tracing of each
individual photon from the exit point on the scintillator surface to
the photosensor, through the plenoptic system.
Thin and thick lenses, together with their apertures, are implemented in the paraxial approximation, assuming 100\% lens transparency and neglecting optical aberrations.
The propagation time of each photon is computed assuming straight-line travel between optical elements and using the speed of light in vacuum.
Further, general surfaces can be added to the
optical stack by defining their shape using tessellated triangles in
an STL file.

The components of the plenoptic camera model include:
the \mla, parametrised as an array of thin lenses or as an array of
pinholes with constant same pitch;
the parameters of the \mla thin-lens model, including the diameter
($D_{\mu}$) and the lateral position ($L_{\mu}$) of each \microlens; and
the distance between the \mla and the sensor planes ($B$) and the
focal length of the \mla ($f_{MLA}$).
The geometry of the \mla accounts for the insensitive areas between
adjacent \microlens{es}.
All the \microlens{es} have the same diameter and focal length.
The \mainlens is modelled with the thick-lens equation, which
includes its diameter ($D_{L}$), the position of the two principal
planes ($P_1$ and $P_2$), and the focal length $f_L$.
This model can be easily converted to that of a thin lens
if the two parallel principal planes coincide.

The simulation of the SPAD array sensors was based on the parameters
of SwissSPAD2~\cite{Ulku_2019_A_512_x}.
The pixel active and dead areas of the SPAD array were simulated as
designed, thus with a geometrical fill factor.
The photon detection efficiency (PDP) was simulated as a binomial
efficiency following the corresponding wavelength spectrum.
The DCR was also cross-checked against measurements performed in our
laboratory, and where explicitly mentioned, extrapolated to a future,
realistic performance.
The timestamp of each pixel can be simulated in either self-trigger
or gate mode.

The positions and the relative distances of the optical objects were
adjusted to reflect the configuration of the PLATON-prototype (see
  Sec.~\ref{sec:results-postprocessing} and
\ref{sec:results-calibration}) or that of an optimised PLATON
detector (see Sec.~\ref{sec:results-platon-nu}).
Although this model appears rather simple, it accurately reproduces
the actual resolution of the PLATON-prototype.

\subsection{Image post-processing method}
\label{sec:results-postprocessing}

The detection of elementary particles in scintillator requires
single-photon detection. Thus, photon-starved images replace the more
traditional ones based on light intensity.
A post-processing method was developed to backtrace all the photon
rays from the image plane (sensor pixels) to the object space (light source).

Post-processing algorithms for plenoptic systems are based on
parallax detection and require pattern matching between images
observed by different \microlens{es}~\cite{konz2016depthest}.
In this reconstruction regime, a single point source would create a sub-image in a set of neighbouring micro-lenses. If the point source sits in the focus region of the plenoptic camera, each microlens images a single pixel, while a point source out of focus would create a blurred image in each participating microlens. The number of active microlenses is known as Virtual Depth (VD) and is proportional to the physical distance of the object from the main lens, allowing for depth estimation.
However, we decided not to adopt this technique as it cannot be
easily applied to photon-starved frames, as it would lead to the
pattern matching failing.
Thus, we require an optical model (see Sec.~\ref{sec:optical-model})
to project each detected photon through the optics into the object space.
Post-processing, based on that of a pinhole camera, is also used for
plenoptic cameras in light-intensity imaging~\cite{chief-ray-hahne}.
In this work, we adopt the same principle to the case of
photon-starved images and, thus, developed the following method: the ray
trace of every photon that was detected by a pixel of the sensor (a
count) is forced to pass through the centre of the closest
\microlens, and interpolated in the object space.
The optical model described in Sec.~\ref{sec:optical-model} was used
for the propagation of the ray tracing:
the \mla was parametrised as an array of pinholes, and each photon
was assumed to have impinged at the centre of the pixel.
Outlier ray traces, mainly generated by noise, but also by the
adopted approximation,
were iteratively rejected until either a predefined minimum number of
rays with the relative closest distance was obtained or the distance
between each leftover ray was below a certain threshold.
The interpolation with the remaining ray traces was used to estimate
the position of the light source.

The constraint on the direction of the photon rays induced by the
pinhole approximation makes this method practical, as the
computational complexity scales linearly with the number of photons detected.
Eventually, the triangulation is performed among the propagated rays
in the object space, based on the minimal distance between different rays.

\subsection{Calibration of the PLATON-prototype}
\label{sec:results-calibration}

The post-processing of the 2D images relies on the accurate
calibration of the optical model of the prototype, that is, the
data-driven inference of the model parameters discussed in
Sec.~\ref{sec:optical-model}.
First, the lateral position of each \microlens is obtained by
illuminating the PLATON-prototype without the \mainlens
with a source of parallel light rays.
This was achieved using a low-power laser mounted on a custom-made
motorised stage, which could be translated parallel to the \mla plane.
A single greyscale intensity image
was obtained from 4'096 frames and used to determine the centre of
each \microlens.
The mean distance between the centre of two adjacent \microlens{es}
was computed to derive the parameter $D_\mu$.
The position of the \microlens{es} could not be determined for the
entire area of the \mla, as the glue partially covers that region
(see Fig.~\ref{fig:platon-prototype}). Consequently, that part of the
image was not used in the post-processing.
This issue reduced the field of view of the PLATON-prototype.
Imaging a parallel-light source is not sufficient to infer the
remaining parameters of the optical model. Hence, we acquired a
dataset consisting of 12-bit images of a back light-illuminated 50
$\mu$m pinhole at 168 different positions within an imaging volume of
$4 \times 4 \times 12~\text{cm}^3$. Each image was obtained by
stacking 4,096 frames, each recorded with a 10~$\mu$s gate.
The motorised stage, by design, allowed the light source to move in
steps of $10~\mu\text{m}$.
Measurements where the light source fell outside the camera's field
of view (FoV), reduced by the glue, were removed from the dataset.
In total, 160 images were used to fit the parameters of the optical
model using a custom genetic algorithm, in which the post-processing
method was iteratively applied and the distance between the
reconstructed and true positions was minimised.

The nominal values of $B$ and $f_L$ provided by the manufacturers
were used as seeds in the fit, while the parameters of the \mla,
previously measured with the parallel-light source, were kept fixed.

The optical simulation (see Sec.~\ref{sec:optical-model}) was used to
reproduce images from a 100~$\mu m$ point source at the same 168
positions as those of the acquired dataset.
The focal length of the \microlens array, $f_{\text{MLA}}$,
was fixed to its nominal value in the datasheet of the \mla.
Then, the simulated images were compared to the corresponding data
samples, and good agreement was observed, indicating that the model
provides a reasonable representation of the camera’s optical stack.

In addition to the calibration sample, an independent
dataset—referred to as the ``validation'' set—was acquired and used
to evaluate the performance of the post-processing method.
In this case, 387 different positions were taken with a depth spacing
of 10 mm over a range of 80 mm.
The lateral spacing was 1 mm, ranging from $-21$ mm to 21 mm, along
one of the lateral axes.
The results of the calibration, as well as the obtained spatial
resolution on light intensity images as a function of the number of
detected photons, are shown in Fig.~\ref{fig:calibration}.
The validation sample exhibits a spatial resolution slightly inferior
to that of the calibration sample. We realised that this was due to a
misalignment caused by the disassembly of the experimental setup
between the acquisition of the two data samples.

\subsection{Data-driven likelihood reconstruction analysis}
\label{sec:likelihood-analysis}

Alternative post-processing methods were also tested.
The first one consists of a data-driven likelihood analysis.
The intensity greyscale images for each of the 160 points scanned
during the calibration campaign were used as probability density
functions (PDFs).
To each PDF, the distribution of the dark counts across the pixels,
obtained with an independent measurement,
is subtracted.
This way we could estimate the intrinsic spatial resolution of the
plenoptic system for the case of a negligible contamination from dark
counts. 
This study aims to emulate the scenario of a future SPAD
array chip capable of rejecting dark counts by time
coincidence by constructing sampled data frames:
after subtracting the independently measured dark-count
map, we generate low-photon-count binary frames by
sampling from the resulting intensity PDFs.
See more
details in the discussion in Sec.~\ref{sec:discussion-conclusions}.
`Sampled'' data frames
were obtained by randomly throwing the position of
the counts
from the PDF of a given position of the light source.

In a given sampled frame, each pixel $i$ can take the value $n_i = 0
\text{ or } 1$, following the Bernoulli probability distribution.
On the other hand, the probability $p_i(\vec{y})$ for a pixel to take
the value of 1 depends also on the position of the point light source
$\vec{y}$ and is described by the corresponding
measured
PDF.
The corresponding negative log-likelihood is given by the binary
cross-entropy (\bce), which compares a sampled data frame taken at
the position $\vec{x}$ to the PDF at a position $\vec{y}$ and is given by
\begin{align}
  \bce(I_{\vec{x}}|\vec{y}) =
  & -\frac{1}{N_{\text{pixels}}}\sum_i^{N_\text{pixels}} n_i
  \log(p_i(\vec{y})) \\
  & + (1-n_i)\log(1-p_i(\vec{y})) \nonumber
\end{align}
where the first term contributes only when the given pixel in the
sampled frame takes the value of 1, while the second term contributes
when the pixel is 0 and weighs the probability of the pixel to be off.
This likelihood definition allows for taking into account both
activated and non-activated pixels.
The prefactor $1/N_{\text{pixels}}$ normalises the \bce\ to an average contribution per pixel, keeping its scale independent of the number of pixels entering the comparison and without affecting the position of the minimum with respect to $\vec{y}$.
For each one, the \bce
was computed with respect to the PDF at every scanned position
$\vec{y}$, and the position yielding the lowest \bce value was
selected as the best-fit estimate.
We refer to this approach as the ``likelihood'' method.

Since the testable resolution of the likelihood method depends on the
distance between two adjacent scanned positions, the calibration
sample was augmented within a smaller portion of the volume, which
was sampled with a higher density of points.
At the depths of 245 mm and 345 mm from the sensor, 101 16-bit images
were taken respectively, with a spacing of 0.1 mm.
These were used to determine the lateral resolution.
For the depth resolution, data points with a spacing of 1 mm along
the optical axis, between a 230 mm and 450 mm distance from the
sensor, were gathered.
Each of these intensity images was used as an empirical
PDF to infer the position of the point light source, and
the corresponding sampled low-photon-count datasets
were generated as described above.

A depth resolution of 2.5~mm was achieved along with a lateral
resolution of 0.1~mm.
With the sampled photon-starved images, the depth and lateral
residual decreased to, respectively, below 10~mm and 0.3~mm for five photons.
Simulation studies have shown that the existence of degeneracies
results in an oscillatory pattern in spatial resolution along both
the lateral and depth directions when fewer than 30 counts are
detected. This effect is due to statistical fluctuations and is not
always reproducible in data, as too many steps of the movement stage
and measurements would be needed.
Below 10 counts, degeneracies arise also from the fact that points
along a single line starting from the principal point of the
\mainlens correspond to concentric images.

\subsubsection{Point-light spatial resolution and comparison with
classical cameras}
\label{sec:results-platon-nu-resolution-point-details-and-classical}

Before explicitly studying the capability of a PLATON-10cm module to
detect GeV neutrinos, a simulation study was performed to evaluate
the spatial resolution to a point light source corresponding to 1 MeV
energy deposition.
The post-processing method used to evaluate the spatial resolution
for a point light source is the same as described in
Sec.~\ref{sec:results-postprocessing}.
For a single-point scintillation light source, creating 10,000
photons, the 3D spatial resolution is 0.3 mm. If the number of
cameras is reduced to two, the 3D spatial resolution becomes 0.8 mm.

For consistency, we compared the performance of the plenoptic cameras
with classical cameras, obtained by removing the \mla and adapting
the lens-to-sensor distance. The post-processing adopted for the
classical cameras is the same as that used for plenoptic cameras,
with the difference that now the pinhole is the \mainlens.
While a single classical camera cannot precisely reconstruct the 3D
position of the light source, the resolution of two orthogonal
cameras is 2.4 mm, which is three times worse than that of the
corresponding plenoptic system.
Moreover, we note that two classical cameras fail to reconstruct the
position of the light source for approximately 23\% of the events,
whereas all events are successfully reconstructed for two plenoptic cameras.
The resolution for eight cameras, split into two orthogonal views, is
approximately four times worse than that of the PLATON-10cm detector
module, which utilises eight plenoptic cameras.

\begin{figure*}[htbp]
  \centering
  \includegraphics[width=0.95\textwidth]{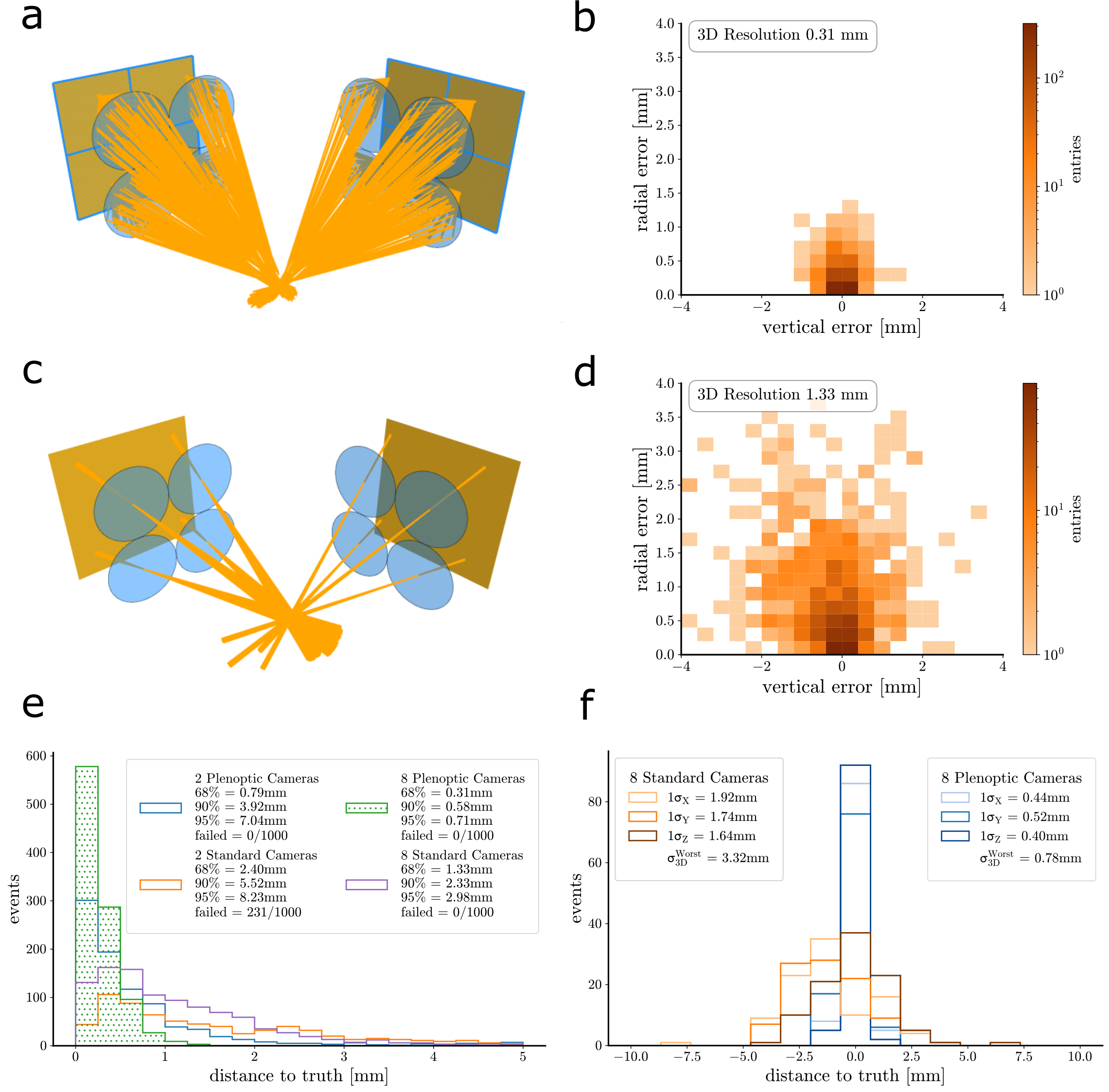}
  \caption{\textbf{Comparison of the performance between plenoptic
    and standard camera versions of the PLATON detector concept.}
    Event reconstruction of a simulated point-like light source with
    eight plenoptic \textbf{a} and classical \textbf{c} cameras.
    The spatial residual of the reconstructed vertical position and
    the cylindrical radial distance from the true origin of photons
    for eight plenoptic in \textbf{b} and standard in \textbf{d} cameras.
    \textbf{e} The distributions of 3D spatial residual, i.e., the
    Euclidean distance between the truth and reconstructed position
    of the source for 1,000 simulated events for eight and two
    orthogonal cameras for plenoptic and classical cameras.
    The legend displays the values of the 68\%, 90\% and 95\%
    percentiles of the one-sided distribution as well as the number
    of events for which the post-processing failed.
    \textbf{f} The distribution of the
    residual of the worst reconstructed 10\% of events is shown for
    both plenoptic (cold-coloured lines) and classical (warm-coloured
    lines) cameras.
    The peeling method, described in
    Sec.~\ref{sec:results-postprocessing}, is not applied to either
    standard or classical cameras.
  }
  \label{fig:CC_PC}
\end{figure*}

Moreover, in the eight-camera setup, 10\% of the events with the
worst resolution show an average residual of 3.3~mm, which is 4.2
times larger than the case of plenoptic cameras (0.8~mm).
This means that plenoptic cameras exhibit a lower probability of
making relatively large mistakes in estimating the light source position.
The results of the simulation are shown in Fig.~\ref{fig:CC_PC}.

Although an array of classical cameras is capable of reconstructing
3D point-like particle interactions in scintillator, the best
performance is provided by an array of plenoptic cameras.
Thus, we opted for this latter configuration as baseline for the
PLATON detector module in this work.

\subsection{Neural-network based reconstruction}
\label{sec:methods-nn}

The dataset used for the neural network studies comprised
1M CC-inclusive muon neutrino interactions generated as described in
Sec.~\ref{sec:results-platon-nu-neutrino}.
The simulation of the detector setup is discussed in
Sec.~\ref{sec:results-platon-nu}.
Reconstructing 3D neutrino interactions within this setup is
inherently challenging. SPAD arrays
are arranged orthogonally on two sides of the scintillator block.
Each interaction produces a photon-starved pattern, typically
comprising between a few tens and several thousand detections across
the SPADs. The task is to infer the spatial origin and topology of
the interaction using only these sparse observations.

To achieve this, we opt for a transformer-encoder-based architecture
to fully exploit the spatial correlations among detected photons
through the transformer's attention
mechanism~\cite{NIPS2017_3f5ee243, NIU202148}. Unlike traditional
architectures that rely on local connectivity or fixed sequence
order, the transformer's attention mechanism enables dynamic
interactions between all elements in the input. This is particularly
beneficial for our application, where the input consists of sparse
and spatially-distributed photon detections and where meaningful
correlations may exist between distant regions of the detector.
Compared to convolutional neural networks (CNNs)~\cite{firstCNN,
9451544} or graph neural nets (GNNs)~\cite{4700287, 9046288,
ZHOU202057}, which are constrained by local or explicitly defined
receptive fields, transformers are not limited to such neighbourhoods
and can integrate global spatial context more effectively. Recurrent
neural networks (RNNs)~\cite{JORDAN1997471, 10.5555/553011,
SHERSTINSKY2020132306} are also ill-suited for this task, as they
assume a consistent sequential ordering, which does not naturally
apply to unordered 2D photon hits~\cite{Xiao2021}. In contrast, the
transformer's attention mechanism provides the flexibility to learn
arbitrary spatial dependencies, which is essential for resolving the
complex and often ambiguous photon distributions resulting from
neutrino interactions~\cite{ISLAM2024122666, 10.1145/3505244}. An
alternative class of models worth considering is Neural Radiance
Fields (NeRFs)~\cite{mildenhall2020nerfrepresentingscenesneural},
which learn continuous volumetric scene representations from dense
multi-view RGB data using ray-based modelling. While NeRFs have
proven effective for reconstructing 3D geometry from 2D projections,
their direct application to our setting is limited. SPAD sensors
produce extremely sparse, binary detections, lacking the dense
photometric supervision and calibrated viewing geometries that NeRFs
rely on. Furthermore, NeRFs are primarily designed for photorealistic
view synthesis, whereas our goal is to recover the underlying spatial
topology of particle interactions from non-image-based, unordered
data. As such, although NeRF-inspired techniques may hold promise for
future work~\cite{yu2021pixelnerfneuralradiancefields,
niemeyer2021regnerfregularizingneuralradiance}, they would require
substantial re-engineering to address the unique challenges of our domain.
Moreover, more traditional reconstruction methods are discarded due
to the multi-dimensional complexity of the problem and the large
number of sparse photons involved. These methods often lack the
flexibility and computational efficiency required to process and
interpret the intricate patterns of photon detection necessary for
accurate 3D reconstruction.

\begin{figure*}[htbp]
  \centering
  \includegraphics[width=1.0\linewidth]{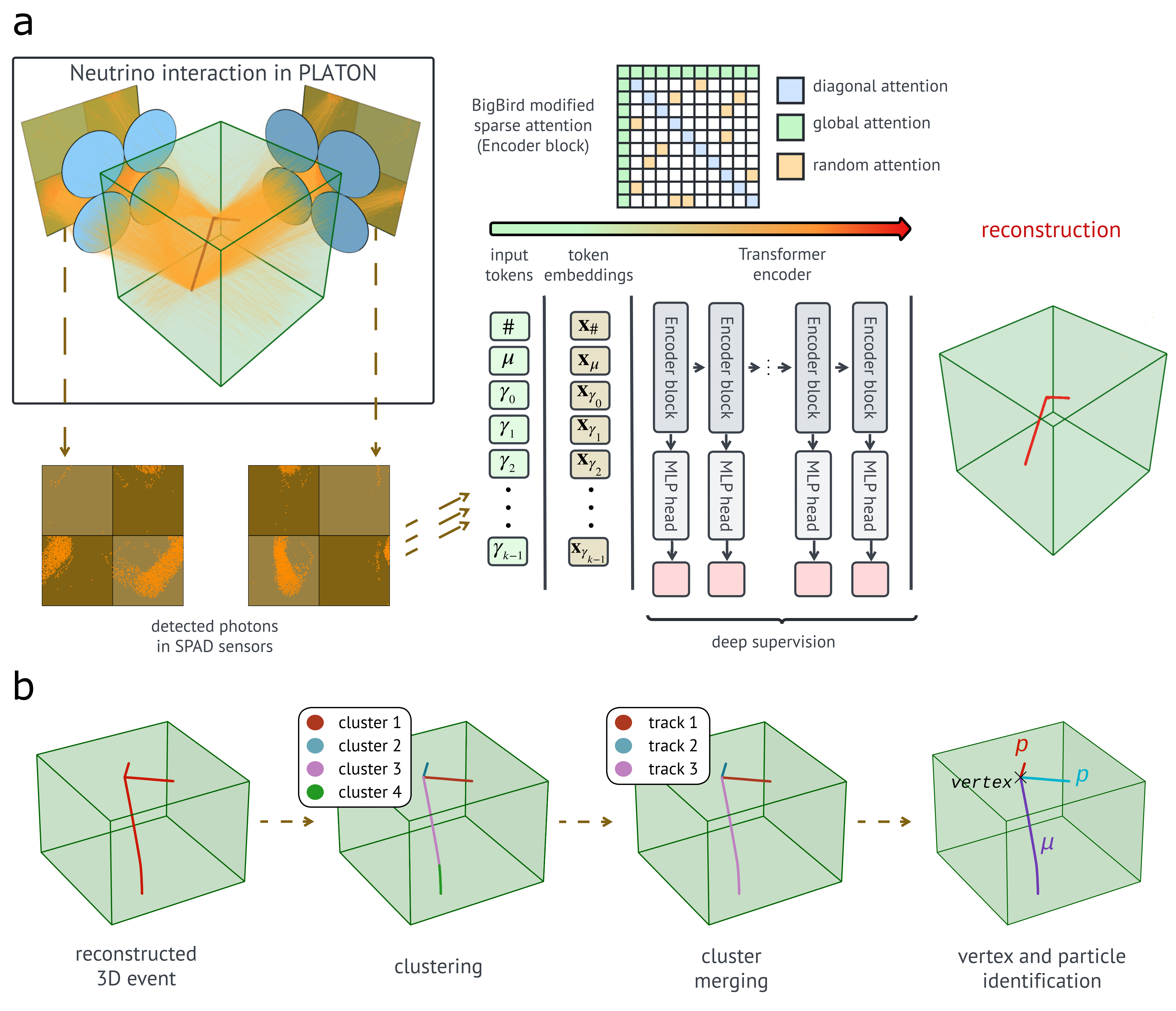}
  \caption{\textbf{Neural event reconstruction and
    pattern-recognition pipeline.} \textbf{a} Neural-network-based
    reconstruction workflow. From left to right: the detected 2D
    photon coordinates are grouped into a sequence of tokens
    $\boldsymbol{\gamma}$, with two additional tokens appended: $\#$,
    indicating the total number of photons, and $\mu$, optionally
    encoding the muon's exiting direction and position. Then, a
    \microlens embedding identifying the optical element that detected
    the photon is summed to the photon coordinate embeddings, and a
    token type embedding (indicating whether the token represents a
    photon coordinate, photon count, or muon exiting information) is
    summed to each token embedding. A modified BigBird Transformer
    encoder~\cite{zaheer2021bigbirdtransformerslonger} processes the
    resulting combined embeddings, which leverages diagonal, global,
    and random sparse attention mechanisms. Finally, multi-layer
    perceptron (MLP) heads with deep supervision are used to perform
    the event reconstruction.
    \textbf{b} Pattern recognition steps. From left to right:
    reconstructed hits are grouped into clusters; collinear and
    nearby clusters are merged together; finally, particle
    identification and vertex reconstruction are performed.
  }
  \label{fig:nn}
\end{figure*}

We employ a transformer encoder to tackle this reconstruction problem
(see Figure~\ref{fig:nn}a). In our setup, each photon is treated as a
token, embedding its 2D position on the SPAD array, the identifier of
the \microlens it passed through, and the corresponding SPAD sensor
ID. Additionally, the input sequence includes a dedicated token
encoding the number of sampled photons in the event and one token for
each muon exiting the scintillator block (typically one), which
contains the 3D exit point and direction vector. The maximum sequence
length is fixed at 1,024 tokens.
During training, we randomly select $\lfloor 0.9 N \rfloor$ of photons per event (with $N$ being the number of photons in the event, further capped by the maximum sequence length) and fill the remainder of the sequence with a padding token, which is ignored during training. This stochastic subsampling means that, from the model's perspective, the same physical event is seen with a different subset of photons at each epoch, and thus acting as an input-level regularisation/data-augmentation mechanism, reducing overfitting to specific photon patterns and improving generalisation to unseen events.
Given that each
token already includes spatial information, explicit positional
encodings are unnecessary. Early experiments showed that the network
struggled to reconstruct events when the number of detected photons
was significantly lower than the maximum sequence length. We
hypothesise that this limitation arises from the model's lack of
awareness of the photon count in each event. To address this, we
include the aforementioned dedicated token encoding the sequence
size, which resolves this issue by enabling the model to condition
the number of observed photons.

The quadratic complexity $O(n^2)$ of the standard transformer's
attention mechanism poses computational challenges for processing
events with a large number of photons. To overcome this, we use a
custom-adapted version of BigBird's sparse
attention~\cite{zaheer2021bigbirdtransformerslonger}, which combines
global, window, and random attention within blocks of tokens to
achieve linear complexity $O(n)$ while effectively modelling
long-range dependencies. Global attention allows specific tokens to
attend to all others, window attention limits tokens to attend only
to their neighbours, and random attention randomly connects tokens.
This hybrid approach allows the model to handle long sequences while
being computationally efficient. In particular, we based our code on
HuggingFace's
implementation~\cite{wolf2020huggingfacestransformersstateoftheartnatural,
bigbird-huggingface}. Specifically, the BigBird configuration used in
our model includes a block size of 64 tokens, with custom
modifications to the attention mechanism. Unlike the original BigBird
implementation, our configuration does not utilise window attention;
each token attends only to itself and not its neighbours (in the
sequence space) or tokens within the same block unless that block is
selected through random attention. Additionally, global attention is
restricted to the global tokens within the first block and does not
extend to all tokens in the block. The random attention spans 4
blocks, a value chosen specifically for this setup. These decisions
in the attention mechanism arise from the fact that we are not
working with true sequential data; the ordering of photons in the
input sequence is arbitrary, and we aim to avoid the model learning
spurious correlations or patterns from this ordering. Our transformer
encoder architecture comprises 12 attention layers, a hidden size of
384 units, GELU
activations~\cite{hendrycks2023gaussianerrorlinearunits}, and 12
attention heads. Shuffling the photon sequence mitigates the risk of
the network learning undesirable block patterns.

The Mean Squared Error (MSE) was our first choice as a loss function.
It minimises the squared difference between the predicted and true 3D
origins of photons. Formally, it is defined as

\begin{equation}
  \mathcal{L}_{\text{MSE}} = \frac{1}{N} \sum_{i=1}^{N} \left\|
  \mathbf{y}_i - \mathbf{\hat{y}}_i \right\|^2,
\end{equation}

where \( \mathbf{y}_i \) and \( \mathbf{\hat{y}}_i \) denote the true
and predicted 3D coordinates of the \( i \)-th photon origin, and \(
N \) is the total number of photons.

Initial experiments, however, revealed convergence difficulties due
to inherent ambiguities in photon patterns. To mitigate this, we
incorporated the Chamfer Distance (CD), commonly used in the point
cloud reconstruction literature. CD relaxes the strict point-to-point
correspondence by minimising the distance between each reconstructed
point and its closest true point (and vice versa) and is given by:

\begin{equation}
  \mathcal{L}_{\text{CD}} = \frac{1}{|\mathbf{\hat{Y}}|}
  \sum_{\mathbf{\hat{y}} \in \mathbf{\hat{Y}}} \min_{\mathbf{y} \in
  \mathbf{Y}} \left\| \mathbf{\hat{y}} - \mathbf{y} \right\|^2
  + \frac{1}{|\mathbf{Y}|} \sum_{\mathbf{y} \in \mathbf{Y}}
  \min_{\mathbf{\hat{y}} \in \mathbf{\hat{Y}}} \left\| \mathbf{y} -
  \mathbf{\hat{y}} \right\|^2.
\end{equation}

where \( \mathbf{Y} \) and \( \mathbf{\hat{Y}} \) are the sets of
true and predicted 3D points, respectively.

We use CD instead of the Earth Mover's Distance (EMD)~\cite{710701}
due to EMD's higher computational complexity and limitations in
preserving the fidelity of detailed
structures~\cite{wu2021densityawarechamferdistancecomprehensive}.
This approach facilitates learning the overall shape of the neutrino
interaction rather than precisely backtracking each individual photon
ray. A known limitation of CD is that it does not account for local
variations in point density. In regions with low ground-truth
density, a single true point may serve as the nearest neighbour for
many predicted points, leading to an over-representation of these
sparse areas in the loss, while regions with denser ground truth are
under-represented. To overcome this limitation, the Density-Aware
Chamfer Distance
(DCD)~\cite{wu2021densityawarechamferdistancecomprehensive} is
adopted. The DCD augments the CD loss by incorporating a
density-aware weighting scheme along with exponential mapping.

For each predicted point \(\hat{y} \in \mathbf{\hat{Y}}\), denote its
nearest neighbour in the true set \( \mathbf{Y} \) and the
corresponding squared distance by
\[
  y^*(\hat{y}) = \arg\min_{y\in \mathbf{Y}} \|\hat{y}-y\|^2, \quad
  d(\hat{y}) = \|\hat{y}-y^*(\hat{y})\|^2.
\]

For each true point \(y \in \mathbf{Y}\), denote its nearest
neighbour in the predicted set \(\mathbf{\hat{Y}}\) and the
corresponding squared distance by
\[
  \hat{y}^*(y) = \arg\min_{\hat{y}\in \mathbf{\hat{Y}}} \|y-\hat{y}\|^2, \quad
  d(y) = \|y-\hat{y}^*(y)\|^2.
\]

In the implementation, the number of times a point is queried is used
for density weighting. Let \(n\bigl(y^*(\hat{y})\bigr)\) be the
number of predicted points that select \(y^*(\hat{y})\) as their
nearest neighbour, and \(n\bigl(\hat{y}^*(y)\bigr)\) be the number of
true points for which \(\hat{y}^*(y)\) is the nearest neighbour. With
a small constant \(\varepsilon\) (i.e., \(1\times10^{-12}\)) and an
exponent \(n_\lambda\) (set to 0.5 in our experiments), the weights are given by

{\small
  \[
    w(\hat{y}) =
    \frac{\gamma_{y\hat{y}}}{\left(n\bigl(y^*(\hat{y})\bigr)^{n_\lambda}
    + \varepsilon\right)}, \quad w(y) =
    \frac{\gamma_{\hat{y}y}}{\left(n\bigl(\hat{y}^*(y)\bigr)^{n_\lambda}
    + \varepsilon\right)}.
  \]
}

Here, the regularisation factors balance the contributions from the
two sets and are defined as
\[
  \gamma_{y\hat{y}} = \frac{n_{y}}{n_{\hat{y}}}, \quad
  \gamma_{\hat{y}y} = \frac{n_{\hat{y}}}{n_{y}}
\]
with \(n_{\hat{y}} = |\mathbf{\hat{Y}}|\) and \(n_y = |\mathbf{Y}|\).

An exponential mapping with scaling factor \(\alpha\) (set to 40 in
our experiments) is applied to the squared distances, so that
\[
  e^{-\alpha\, d(\hat{y})} \quad \text{and} \quad e^{-\alpha\, d(y)}
\]
map each squared distance to the interval \((0,1)\). Consequently,
the Density-Aware Chamfer Distance loss is expressed as

\begin{align}
  \mathcal{L}_{\text{DCD}} &= \frac{1}{2} \Bigg(
    \frac{1}{|\mathbf{\hat{Y}}|} \sum_{\hat{y}\in \mathbf{\hat{Y}}}
    \left( 1 - w(\hat{y})\, e^{-\alpha\, d(\hat{y})} \right) \notag \\
    &\qquad\quad + \frac{1}{|\mathbf{Y}|} \sum_{y\in \mathbf{Y}}
    \left( 1 - w(y)\, e^{-\alpha\, d(y)} \right)
  \Bigg).
\end{align}

For small distances, the contribution is approximately linear; for
larger distances, the contribution saturates, thereby reducing the
influence of outliers. The total loss applied to the transformer
encoder layers comprises both the DCD and a weighted MSE loss,
ensuring a balanced learning process. Deep
supervision~\cite{lee2014deeplysupervisednets,
li2022comprehensivereviewdeepsupervision} is applied by computing the
loss at each encoder layer with exponentially increasing weights. For
a transformer with \( l \) encoder layers, the total loss is

\begin{equation}
  \mathcal{L}_{\text{total}} = \sum_{i=1}^{l} \frac{1}{2^{(l-i)}}
  \left( \mathcal{L}_{\text{DCD}} + \beta\, \mathcal{L}_{\text{MSE}} \right),
\end{equation}

where \( \beta \) is a scalar weight that balances the MSE loss with
the DCD loss (set to 1 in our experiments), and the factor \(
\frac{1}{2^{(l-i)}} \) assigns greater weight to losses computed at
deeper layers.

This combined loss function enables learning both the overall
geometry and the fine-grained spatial photon distribution
effectively, leading to improved convergence and reconstruction quality.

The model was implemented in Python~3.12.2 using
PyTorch~2.5.1~\cite{NEURIPS2019_9015} and PyTorch
Lightning~2.4.0~\cite{Falcon_PyTorch_Lightning_2019}, and trained on
a single NVIDIA H100 GPU with 94~GB of memory. The training was
conducted over 200 epochs with a batch size or 64 and a maximum input
sequence length capped at 1024, as previously specified. The
optimiser employed was AdamW~\cite{Kingma-2014-adam,
loshchilov2019decoupledweightdecayregularization}, with a learning
rate initialised at \(2 \times 10^{-4}\), \(\beta_1 = 0.9\),
\(\beta_2 = 0.95\), and a weight decay of \(10^{-4}\). The learning
rate schedule consisted of a single warm-up epoch, followed by cosine
annealing over the remaining 199 epochs, decaying smoothly to zero.
Hyperparameter optimisation was conducted using the Optuna framework,
enabling efficient search space exploration through automated
sampling and pruning strategies~\cite{optuna_2019}. No dropout was
applied, as empirical evidence indicated a detrimental effect on the
regression of precise continuous values. The generalisation capacity
typically provided by dropout was instead achieved through the
stochastic sampling of photons per event. Given the large number of
possible photon combinations, the likelihood of the model
encountering the same subset of inputs across epochs is negligible. All results presented in Sec.~\ref{sec:results-platon-nu-neutrino} were obtained on the held-out test set from a $60\%/10\%/30\%$ train/validation/test split, respectively.

\subsection{Details of the neutrino simulation analysis}
\label{sec:results-platon-nu-neutrino-details}

\subsubsection{Post-processing}
\label{sec:results-platon-nu-neutrino-details-postprocessing}

The neutrino event post-processing consists of the following steps:
first, a significant fraction of photosensor dark counts are rejected
by signal coincidence in a time window of 10 ns;
the 2D images collected by the eight cameras and the muon smeared
exiting position (mean = 0 mm, std = 1 mm)
and direction (mean = 0 radians, std = 0.05 radians)
are given as input to the neural network (NN). No timing information
is passed to the NN, which instead returns the 3D origin of each
photon associated with a count in the scintillator volume.
The result is a cloud of points (``\photorig{s}'') that indicate the
origin of the scintillation photons along the trajectory of every
charged particle.
This is illustrated in Fig.~\ref{fig:neutrino-postprocessing}.

\subsubsection{Pattern recognition}
\label{sec:results-platon-nu-neutrino-details-pattern-recognition}

The reconstruction of PLATON 3D images differs from what data
processing in particle physics experiments typically requires, and is
illustrated in Fig.~\ref{fig:nn}b.
The clustering algorithm is based on Gaussian Mixture Model
(GMM)~\cite{10.1111/j.2517-6161.1977.tb01600.x}:
a probabilistic model that represents complex data as a combination
of simpler sub-distributions, Gaussian in this case, of underlying
hidden classes.
The optimal number of clusters was selected using the elbow method, a
heuristic that identifies the point beyond which increasing the
number of clusters yields only marginal improvements in model performance.
In a second step, principal component analysis (PCA) was used to
merge clusters into tracks, each assigned to a different particle.
This process iterated through all pairs of clusters, merging two
clusters if they were both collinear and sufficiently close to each other.
Once all the particle tracks of the event were obtained, the vertex
of the neutrino interaction was reconstructed.
Since the muon is assumed to be determined by an external detector,
we use its exiting point on the scintillator surface to assign one of
the tracks to the muon.
Finally, the extreme of the muon track opposite to its exiting point
is identified as the neutrino interaction vertex.
In some cases, multiple scattering can break the muon track into
different consecutive segments. To overcome this issue, the
clustering algorithm is rerun with slightly relaxed conditions,
potentially adding a new cluster to the muon track, until a connected
but non-collinear cluster is found.
If the muon does not escape the scintillator volume, the longest
reconstructed track is assigned to it, and the end with the largest
number of connected clusters is identified as the interaction vertex.
If the two track ends exhibit an equal number of connected clusters,
the point forming the smaller angle with the connected cluster is
selected as the interaction vertex.

\subsubsection{Selection of neutrino events}
\label{sec:results-platon-nu-neutrino-details-selection}

The \ccOneMuZeroPiOneProt sample is the most common final-state
topology below 1 GeV, observed in experiments such as T2K or
Hyper-Kamiokande, since it is dominated by CC quasi-elastic (CCQE)
interactions. Another type of final-state topology is named 2 protons
- 2 holes (2p2h) and can be induced by three main types of processes:
meson exchange current, where the momentum transfer is shared between
two nucleons via the exchange of a virtual meson~\cite{ALVAREZRUSO20181};
short-range nucleon-nucleon correlations
(SRC)~\cite{RevModPhys.89.045002}, where the neutrino interacts with
a nucleon that is part of a correlated nucleon-nucleon pair and both
nucleons are knocked out of the nucleus;
FSI, with a nucleon that knocks out a second nucleon.
2p2h events are less frequent ($\sim 8 \%$) than CCQE interactions
($\sim 43\%$), but are notoriously affected by significant systematic
uncertainties in the modelling of the
process~\cite{Gran:2013kda,Nieves:2012yz,Sobczyk:2020dkn}. Thus, the
correct modelling of 2p2h interactions is of vital importance for the
ongoing precision measurements of neutrino
oscillations~\cite{2p2h-nova,PhysRevD.105.072001,PhysRevD.96.092006,t2k-oa-2p2h-2}
and, in particular, for the next-generation LBL experiments that will
search for leptonic charge-parity
violation~\cite{Abe:2018uyc,Abi:2020qib,DUNE:2021mtg}.

The accurate identification of the type of each particle produced by
the neutrino interaction relies on the precise measurement of the
energy loss along its trajectory.
For instance, the detection of the Bragg peak of particles stopping
in the scintillator ensures unambiguous particle identification
(PID). This feature is enhanced by the fact that most of the muons
are produced with a momentum peaked around 0.5 GeV/c with a long tail
above 1 GeV, while protons are expected dominantly below 1 GeV/c with
a peak around 300 MeV/c.
The reconstructed distribution of the energy loss along the track of
protons (often stopping in the scintillator) and muons is shown in
Fig.~\ref{fig:neutrino-pattern-recognition}.
To identify the particle type associated with each reconstructed
track independently, a boosted decision tree (BDT) classifier was
employed using the XGBoost algorithm~\cite{10.1145/2939672.2939785}.
The kinematic variables used as input features to the BDT include:
the variance of the number of photons per mm,
the total number of photons in the track,
the maximum number of photons per mm,
the minimum number of photons per mm,
and the mean number of photons per mm.

\subsubsection{Selection of neutrino events with a system of classical cameras}
\label{sec:results-platon-nu-neutrino-details-selection-classical}
The same study has been conducted with a configuration of the PLATON-10cm
module where a conventional optical system is obtained by removing the \mla.
It was found that, although a system based on standard cameras
exhibits inferior spatial resolution compared to an equivalent system
employing plenoptic cameras (see
Sec.~\ref{sec:results-platon-nu-resolution-point-details-and-classical}),
even orthogonal arrays of conventional cameras instrumented with SPAD
array sensors can achieve very good performance within a PLATON-10cm module over a range of 10~cm.
On the other hand, the drop in selection efficiency below 50\% occurs
at 215~MeV/$c$ (corresponding to a 3.5~mm proton range) with the
\mla, and at 230~MeV/$c$ (corresponding to a 5~mm proton range)
without the \mla.
A similar difference can be found in the neutrino vertex resolution
and the proton angular resolution: the vertex resolution drops from
0.4~mm to 0.6~mm, while the angular resolution falls from
$1.5^{\circ}$ to $1.8^{\circ}$.
This is in line with the results shown in
Sec.~\ref{sec:results-platon-nu-resolution-point-details-and-classical}.
Without the \mla, the efficiency and the purity are, respectively,
6\% and 5\% lower compared to the plenoptic configuration.
It is worth noting that the PLATON-10cm module is required to cover a
depth of field of 10~cm; therefore, it is plausible that eight
classical cameras combined can achieve a reasonably good depth resolution.

\bibliography{sn-bibliography}

\backmatter

\section*{Acknowledgements}

This work was supported by the Swiss National Science Foundation
under grant  PCEFP2\textunderscore203261. This research was also
partially supported by the Swiss National Science Foundation (grant
  20QT21\textunderscore187716 Qu3D ``Quantum 3D Imaging at high speed
and high resolution").
Neural network training in this work used resources of the National Energy Research Scientific Computing Center (NERSC), a U.S. Department of Energy User Facility, under the AI4Sci@NERSC award DDR-ERCAP0034642. Additional training support was provided through the Swiss AI Initiative via a grant from the Swiss National Supercomputing Centre (CSCS), project ID a149, on the Alps system.
We would like to thank Prof. Andr\'e Rubbia from ETH Zurich for useful inputs and discussions and for providing access to laboratory equipment and facilities; 
Prof. Vincenzo Berardi at Politecnico di Bari, Dr. Umut Kose and Johannes W\"uthrich from ETH Zurich for useful discussions; 
Arne Erdmann from Raytrix GmbH for helping to understand the functioning of the plenoptic camera prototype assembled at Raytrix and the use of the RxLive software;
Prof.~Wallny at ETH Zurich for providing access to his group's thermal chamber.

\section*{Author contributions}

D.S. conceived the PLATON detector, is the PI of the project funded
by the Swiss National Science Foundation, and supervised every aspect
of the project.
T.D. was the main analyser and developer.
The plenoptic system of the prototype was designed and built by Raytrix GmbH.
E.C., C.B., and K.K. provided the SPAD array photosensor, assisted
with its use, offered guidance, and supervised the project to ensure
understanding of the results.
T.D., T.W., and M.F. set up the tests in the laboratory.
T.D. ran the experiments, developed the software and analysed the data.
T.D., S.A.-M., and D.S. conceived the standard image post-processing
method, and T.D. developed and tested it.
T.D. developed and tested the software for the simulation of the detector.
T.D. studied the detector configuration of the simulated physics experiments.
Also, C.A. worked on the detector simulation.
S.A.-M. conceived, developed, trained, and validated the
neural-network-based image post-processing.
N.B., S.A.-M., and T.D. performed the pattern recognition and data
analysis of the simulated neutrino experiment.
All the authors contributed to the writing of the paper.

\section*{Disclaimer notice}

This document was prepared by Swiss Federal Institute of
Technology-Zurich (ETH Zurich), in part as a result of the use of
facilities of the U.S. Department of Energy (DOE), which are managed
by The Regents of the University of California, acting under Contract
No. DE-AC02-05CH11231. Neither The Regents of the University of
California, DOE, the U.S. Government, nor any person acting on their
behalf: (a) make any warranty or representation, express or implied,
with respect to the information contained in this document; or (b)
assume any liabilities with respect to the use of, or damages
resulting from the use of any information contained in the document.

\end{document}